\documentclass{cernrep} 
\usepackage{texnames}
\usepackage[T1]{fontenc}
\usepackage[bookmarks, colorlinks=true, linktoc=page, pdftex, linkcolor=black, citecolor=black, urlcolor=blue]{hyperref}
\usepackage{amsmath} 
\usepackage{rotating}
\DeclareMathOperator{\Grad}{\nabla}
\DeclareMathOperator{\Div}{\nabla \cdot}
\DeclareMathOperator{\Curl}{\nabla \times}
\renewcommand{\vec}[1]{\mathbf{#1}} 
\newcommand{\me}[1]{\mathrm{e}^{#1}}
\newcommand{\parder}[2]{\frac{\partial {#1}}{\partial {#2}}}
\newcommand{\pardoubleder}[2]{\frac{\partial^2 {#1}}{\partial {#2^2}}}

\usepackage{fancyhdr}
\fancyhfoffset{4 mm}
\fancypagestyle{ARTTITLE}{%
\fancyhf{} 
\lhead{\small{Proceedings of the 2018 CERN--Accelerator--School course on \it{Beam Instrumentation}, Tuusula, (Finland)}} 
\lfoot{Available online at \url{https://cas.web.cern.ch/previous-schools}}
\rfoot{\thepage\hspace*{3mm}}
 
}
\begin{document}
\title{Introduction to Optics and Lasers for Beam Instrumentation\\
}
 
\author {Stephen Gibson}

\institute{John Adams Institute for Accelerator Science, Royal Holloway University of London, United Kingdom.}

\begin{abstract}
The versatility of optics enables the design of a wide range of elegant beam instrumentation. Multiple properties of particle beams can be precisely measured by various optical techniques, which include:  direct sampling of optical radiation emitted from a charged particle beam;  monitoring interactions with an optical probe such as a laserwire;  and by electro-optic conversion of the beam signal with high-bandwidth fibre readout. Such methods are typically minimally-invasive and non-destructive, thus permitting diagnostics during accelerator operation without perturbation of the particle beam or risk of damage to the instrument. These proceedings summarise three CAS lectures that introduce the basic principles of optics relevant for instrumentation design, outline the key laser technologies and setups, and review the state-of-the-art in laser-based beam instrumentation.
\end{abstract}


\maketitle 
\thispagestyle{ARTTITLE}
%
 
\section{Introduction to Optics: basics, components, diffraction }
\subsection{Motivation}
\label{sec:opticsintro}
\emph{Optics} concerns the behaviour and properties of light, including the transmission and deflection of radiation, whilst \emph{photonics} is the science and technology of controlling and detecting photons. An understanding of both is essential when designing optical beam instrumentation. These lectures aim to equip the reader with enough knowledge of optics, lasers and practical setups to understand and start to develop versatile and precise beam diagnostics.

An increasing variety of beam instrumentation relies on optical techniques to precisely measure the parameters of particle beams. The manner in which the detected photons are produced enables some broad categories of optical beam instrumentation to be identified:

\begin{description}
\item [a) Synchrotron radiation emitted from charge particles] The bremsstrahlung radiation that results when a charged particle is deflected in a magnetic field can be directly measured by downstream photodetectors, typically for beam profile and halo monitoring.
\item [b) Primary beam interactions with laser generated photons] Injecting a tightly focused laser beam into the beampipe facilitates direct interactions between the photons and charged particles, typically via Compton scattering or the excitation of atomic electrons in the case of ions. Examples include laserwires, Shintake monitors and laser polarimeters that use laser beams to probe and measure beam profile, emittance and polarization.
\item [c) Secondary photons from beam interactions] Charged particles interacting with scintillator screens, or fluorescence of residual gas interactions, can produce secondary photons that can be viewed by cameras placed outside the beam pipe and are typically useful for precise beam profile monitoring. 
\item [d) Beam-field induced optical radiation] Direct measurements of Cherenkov, Smith-Purcell, or Optical Transition Radiation, arising from the electromagnetic field of the charged particle beam interacting with a nearby dielectric target.
\item [e) Electro-optic conversion of the beam signal] Replicating the beam signal on an externally generated laser pulse by electro-optic sampling, typically with high-bandwidth readout, for longitudinal bunch profiling.
\end{description}

The main remit for these lectures was to focus on non-invasive, laser-based applications in beam instrumentation [items (b) and (e)], however, the fundamental principles are also pertinent to direct observations of synchrotron radiation, Cherenkov radiation, Smith-Purcell, and Transition Radiation, and to the design of optical arrangements to view invasive diagnostics, such as scintillation screens. In all such instruments, care must be taken to reduce optical aberrations attributed to geometrical refraction as well as interference and diffraction effects. An introduction to the methods to calculate such effects will be provided. This introductory section begins with an overview of basic optics principles, before identifying the main laser technologies and finally reviews a selection of state-of-the-art examples in laser-based beam instrumentation.

\subsection{Geometric Optics}
\subsubsection{Solutions to Maxwell's equations}
Starting from James Clerk Maxwell's equations (1865) for electric $\vec{E}$ and magnetic $\vec{B}$ fields in vacuum in the absence of charge ($\rho=0$) and current ($\vec{J} = 0$),
\begin{align}
&\text{Gauss's law for the electric field:} &\Div\vec{E} &= \frac{\rho}{\epsilon_0} =  0\label{eqn:Gauss} \\
&\text{No magnetic monopoles:} &\Div\vec{B} &= 0\label{eqn:divB} \\
&\text{Faraday's law of induction:} &\Curl\vec{E} &=-\parder{\vec{B}}{t}\label{eqn:Faraday} \\
&\text{Amp\`{e}re's law:} &\Curl\vec{B} &=\mu_0  \vec{J}+ \epsilon_0 \mu_0\parder{\vec{E}}{t} = \epsilon_0 \mu_0 \parder{\vec{E}}{t}, \label{eqn:Ampere}
\end{align}
we take the curl of Faraday's law (Eq.~(\ref{eqn:Faraday})) and apply the vector identity on the LHS: $\Curl (\Curl \vec{A}) = \Grad (\Div \vec{A}) - \nabla^2\vec{A}$,
\begin{align}
\Curl ( \Curl\vec{E} ) &=-\parder{(\Curl\vec{B})}{t} \, \\
\Grad (\Div \vec{E}) - \Grad^2\vec{E} &=-\parder{(\Curl\vec{B})}{t} .
\end{align}
Note from Gauss's law (Eq.~(\ref{eqn:Gauss})) that $\Div \vec{E} = 0$, and applying Amp\`{e}re's law (Eq.~(\ref{eqn:Ampere})) on the RHS to derive for $\vec{E}$,
\begin{align}
\Grad^2\vec{E} &=\parder{(\Curl\vec{B})}{t} \,\\
\Grad^2\vec{E} &=\epsilon_0 \mu_0 \pardoubleder{\vec{E}}{t}.
\end{align}
Similarly, by taking the curl of Amp\`{e}re's law (Eq.~(\ref{eqn:Ampere})), applying the same vector identity on the LHS, noting Eq.~(\ref{eqn:divB}), and applying Eq.~(\ref{eqn:Faraday}) on the RHS we derive for $\vec{B}$,
\begin{align}
\Curl ( \Curl\vec{B} ) &=-\epsilon_0 \mu_0 \parder{(\Curl\vec{E})}{t}\\
\Grad (\Div \vec{B}) - \Grad^2\vec{B} &=-\epsilon_0 \mu_0\parder{(\Curl\vec{E})}{t}\\
\Grad^2\vec{B} &=\epsilon_0 \mu_0\parder{(\Curl\vec{E})}{t}\\
\Grad^2\vec{B} &=\epsilon_0 \mu_0 \pardoubleder{\vec{B}}{t}.
\end{align}
We identify wave equations of the form
\begin{equation}
\nabla^2\psi = \frac{1}{v^2}\pardoubleder{\psi}{t}\,,
\end{equation}
for $\vec{E}$ and $\vec{B}$, and find that $ v = \frac{1}{\sqrt{\epsilon_0 \mu_0}} = 299,792,458$\,m\,s$^{-1}$ ; the electromagnetic wave travels at the~speed of light in vacuum. In three dimensions, one solution to these wave equations is plane waves, 
\begin{equation}
U(x,y,z,t) = U_0 e^{i(\vec{k}\cdot\vec{r} - \omega t)}\, ,
\end{equation}
where $\vec{k}$ is the wave vector, $\vec{r}$ is the position and the angular frequency is $\omega = 2\pi \nu$.

Thus from the plane wave solution we infer that in an isotropic medium light travels in straight lines known as rays, with wavefronts of constant phase orthogonal to the propagation direction. The~phase difference between multiple optical paths will be essential when describing interference and diffraction effects, as we shall in later sections. However, we first briefly review \emph{Geometric Optics}, which is the technique for determining the light path through multiple interfaces between media of different refractive indices. 

\subsubsection{Basics of refractive systems}
\begin{figure}[h]
\begin{center}
\includegraphics[height=30mm]{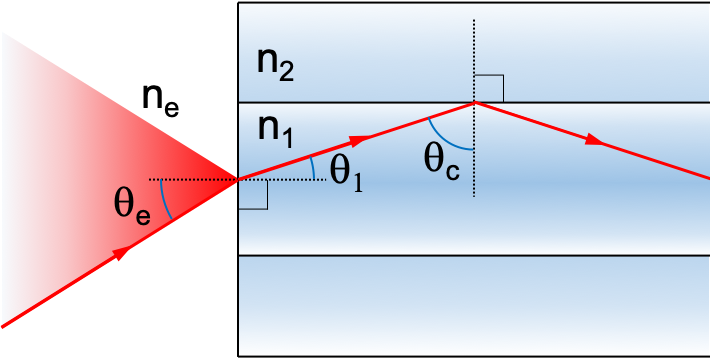} 
\caption{Numerical aperture of a step-index optical fibre}
\label{fibreinputfigure}
\end{center}
\end{figure}

\emph{Geometric Optics} is often sufficient for building basic optical systems in beam instrumentation, and is based on two simple assumptions that are valid for isotropic media and for apertures much larger than the wavelength of light:
\begin{enumerate}
\item light travels in straight lines, known as rays, in a medium of uniform refractive index, $n$.
\item light reflects and/or refracts at an interface between media of different refractive indices.
\end{enumerate}
It is well known that light travels at different speeds in each medium according to $v = c/n$, so a ray of light incident at angle $\theta_1$ to the normal of an interface between two media of difference refractive indices emerges at a refraction angle $\theta_2$, according to Snell's law $n_1 \sin \theta_1 = n_2 \sin \theta_2$. The critical angle, $\theta_c = n_2/n_1$, is the angle of incidence beyond which light passing from a high to low refractive index medium is totally internally reflected. These simple rules are useful in determining the properties of various optical systems. For example, in a step-index optical fibre as shown in Fig.~\ref{fibreinputfigure}, the numerical aperture that determines the acceptance cone or angular spread of emitted radiation can be derived in terms of the refractive indices of the fibre core, $n_{\rm 1}$, and fibre cladding, $n_{\rm 2}$, where $n_{\rm e}$ is the refractive index of the external medium:
\begin{equation}
{\rm N.A.} = n_{\rm e} \sin{\theta_{\rm e}} = \sqrt{n_1^2 - n_2^2} \,.
\end{equation}

Similarly, the application of Snell's law at a double spherical refractive interface results in the~thin lens equation that relates the (negative) object distance, $x_1$, and image distance, $x_2$, from the line of action of a lens, to the focal length, $f$, taken to be positive for a converging lens or negative for a~diverging lens. The expressions are valid in the paraxial approximation for rays that lie close to the optical axis.
\begin{equation}
{\rm \mbox{Lens equation,}}\: \frac{1}{x_2} - \frac{1}{x_1}  = \frac{1}{f}\,;\quad{\rm Lateral\:magnification,}\:m = \frac{h_2}{h_1}  = \frac{x_2}{x_1}\,.
\end{equation}
Further details on the applications of lens equations are described in numerous textbooks, e.g.~\cite{bib:Longhurst, bib:JenkinsWhite, bib:Fowles, bib:SmithKing, bib:Hecht}.

\subsubsection{Instrument design and ray tracing}
Constructing an optical instrument typically requires multiple lenses. If only a small number of lenses is required, then the lens equation may be applied multiple times, or the effective focal length of a~combination of lenses can be found using, e.g.: 
\begin{align}
{\rm \mbox{Two thin lenses in contact,}}\: \frac{1}{f_C} &= \frac{1}{f_A} + \frac{1}{f_B} = \frac{1}{x_2} - \frac{1}{x_1} \,;\\
{\rm \mbox{or when separated by distance} }\,d,\:\frac{1}{f_C} &= \frac{1}{f_A} + \frac{1}{f_B} - \frac{d}{f_A f_B}\,.
\end{align}
However, for longer sequences of lenses, a transfer matrix may be assigned to each component so that rays can be traced through the optical system numerically. Such a method is analogous to particle tracking through an accelerator lattice, though note that an optical converging lens focuses in both planes simultaneously (unlike a quadrupole magnet). Each ray is described by the initial height, $h_{1}$, and initial propagation angle $h_{1}^{'}$, with respect to the optical axis, and transfer matrices are defined as follows:
\begin{equation}
\mbox{Free space drift:}
\begin{cases}
  h_{2} = h_{1} + L h_{1}^{'}\\
  h_{2}^{'} = h_{1}^{'}
  \end{cases}
\implies
\left ( \begin{array}{c}
  h_{2}\\
  h_{2}^{'}
  \end{array}
\right )
 = 
\left ( \begin{array}{cc}
  1 & L\\
  0 & 1
  \end{array}
\right )
\left ( \begin{array}{c}
  h_{1}\\
  h_{1}^{'}
  \end{array}
\right )
\end{equation}
\begin{equation}
\mbox{Action at a thin lens:}
\begin{cases}
  h_{2} = h_{1}\\
  h_{2}^{'} = h_{1}^{'} - h_{1}/f
  \end{cases}
\implies
\left ( \begin{array}{c}
  h_{2}\\
  h_{2}^{'}
  \end{array}
\right )
 = 
\left ( \begin{array}{cc}
  1 & 0\\
  -1/f & 1
  \end{array}
\right )
\left ( \begin{array}{c}
  h_{1}\\
  h_{1}^{'}
  \end{array}
\right )
\end{equation}

\noindent A ray traced through an entire optical sequence is found simply by multiplying all matrix elements, in the~order that the light passes through them (with the first on the right).
For example, for a simple sequence of drift of length $L_1$, followed by a lens of focal length $f$, then a second drift of length $L_2$, the~transfer matrix is:
\begin{equation}
M = 
\left ( \begin{array}{cc}
  1 & L_2\\
  0 & 1
  \end{array}
\right )
\left ( \begin{array}{cc}
  1 & 0\\
  -1/f & 1
  \end{array}
\right )
\left ( \begin{array}{cc}
  1 & L_{1} \\
  0 & 1
  \end{array}
\right )
=
\left ( \begin{array}{cc}
  1-\frac{L_2}{f} & L_{1} - \frac{L_1 L_2}{f} + L_2\\
  -\frac{1}{f} & -{L_1}{f} + 1
  \end{array}
\right )
\end{equation}

When designing an optical instrument the response to the full input light field at various wavelengths is typically required. Ray tracing software divides the real light field into discrete monochromatic rays that are propagated through the optical system.

\begin{figure}[h]
\begin{center}
\includegraphics[width=0.33\textwidth]{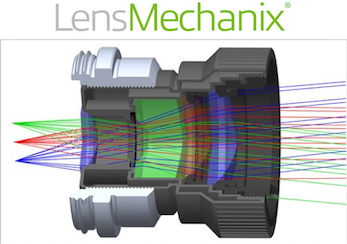}
\includegraphics[width=0.33\textwidth]{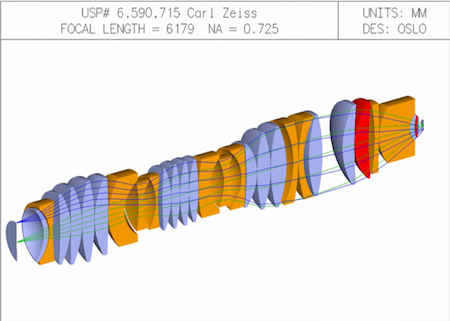} 
\includegraphics[width=0.32\textwidth]{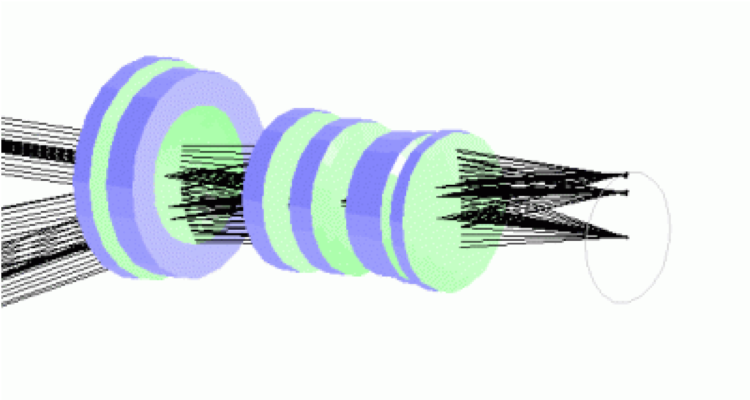} 
\caption{Examples of optical ray tracing software: Zemax\cite{bib:Zemax}, OSLO~\cite{bib:OSLO} and WinLens3D~\cite{bib:WinLens}}
\label{fig:raytracingsoftware}
\end{center}
\end{figure}

 Several professional software suites are available with various utilities, including the ability to input a real light distribution. A selection of professional optical software tools include, Zemax~\cite{bib:Zemax}, OSLO~\cite{bib:OSLO}, and WinLens3D~\cite{bib:WinLens}, as in Fig.~\ref{fig:raytracingsoftware}, and there are open source examples to explore~\cite{bib:RayOpticsSimulation}.

Such optical software enables common aberrations to be assessed and mitigated prior to building the instrument, for example:
\begin{description}
\item[Spherical aberration] Rays striking a lens or mirror off-axis fail to converge at the focus (e.g. optical surfaces are often spherical, so are further from parabolic at larger radii) 
\item[Comatic aberration] Wavefront distortions appear for object points off-axis with a comet-like spread.
\item[Chromatic aberration] Light of different colours refractions through different angles due to dispersion (typically corrected with an achromatic doublet).
\item[Astigmatism] A cylindrical wavefront aberration creating a focus shorter in one plan than in the orthogonal plane.
\end{description}
Apertures stops may be added to limit aberrations at a lens by masking off-axis components. 

\subsubsection{Scintillation screens and the Scheimpflug principle}
Before moving on from \emph{Geometric Optics}, it is worth mentioning one neat and effective optical arrangement often applied in beam instrumentation when viewing scintillator screens. Precise measurements of the size, profile and position of a particle beam striking a scintillator screen requires a carefully designed optical system to transfer scintillation light to the camera, so that the true particle distribution can be reconstructed. The challenge for the optical system is to capture a clean, sharply focused image of the~scintillation plane, free of distortion, optical aberrations, non-linearity, or optical backgrounds (OTR). The image must remain in focus across the whole field-of-view, despite the typically small depth-of-field of a camera placed outside the beampipe and viewing the screen obliquely through a vacuum viewport.
\begin{figure}[h]
\begin{center}
\includegraphics[width=0.99\textwidth]{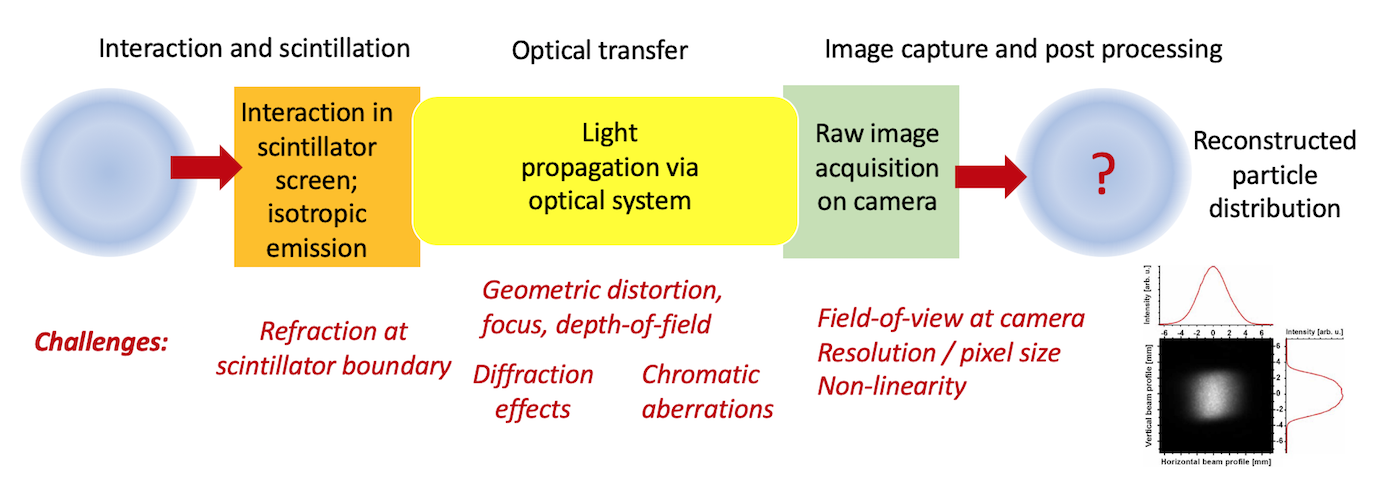}
\caption{Challenges when imaging a particle beam distribution at a scintillation screen}
\label{fig:scintillation}
\end{center}
\end{figure}

\begin{figure}[h]
\begin{center}
\includegraphics[width=0.75\textwidth]{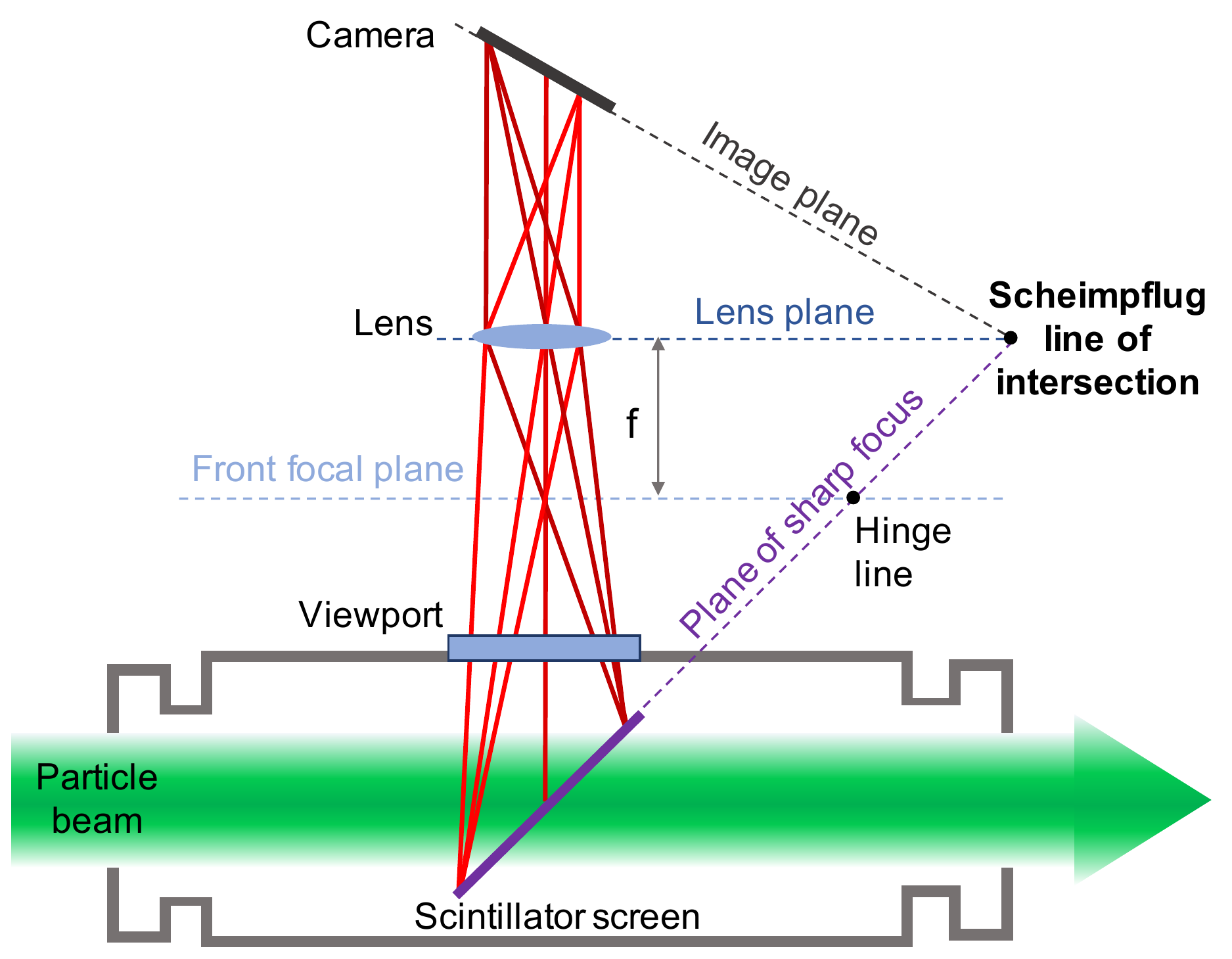}
\caption{The Scheimpflug principle applied to view a scintillator screen}
\label{fig:Scheimpflug}
\end{center}
\end{figure}

Captain Theodor Scheimpflug was an Austrian Army Navel officer who used aerial photography to make accurate maps with undistorted images from balloon-suspended cameras (not pointing vertically down). The Scheimpflug principle states that when the subject plane, lens plane and image plane intersect in a single line, then the subject plane is completely in sharp focus~\cite{bib:Scheimpflug1904}. So by arranging for an appropriate tilt angle between the scintillator screen, an intermediate lens, and the camera plane, the particle distribution at  the scintillator screen can be accurately imaged. An example Scheimpflug arrangement is depicted in Fig.~\ref{fig:Scheimpflug}, in which the scintillator screen is tilted at 45$^\circ$ with respect to the incident particle beam. Provided that the above Scheimpflug condition is met, then the lens forms a sharp, inverted image of the whole plane of the scintillator screen on the camera plane, thus the full particle beam distribution can be captured in focus.

\begin{figure}[h]
\begin{center}
\includegraphics[width=0.99\textwidth]{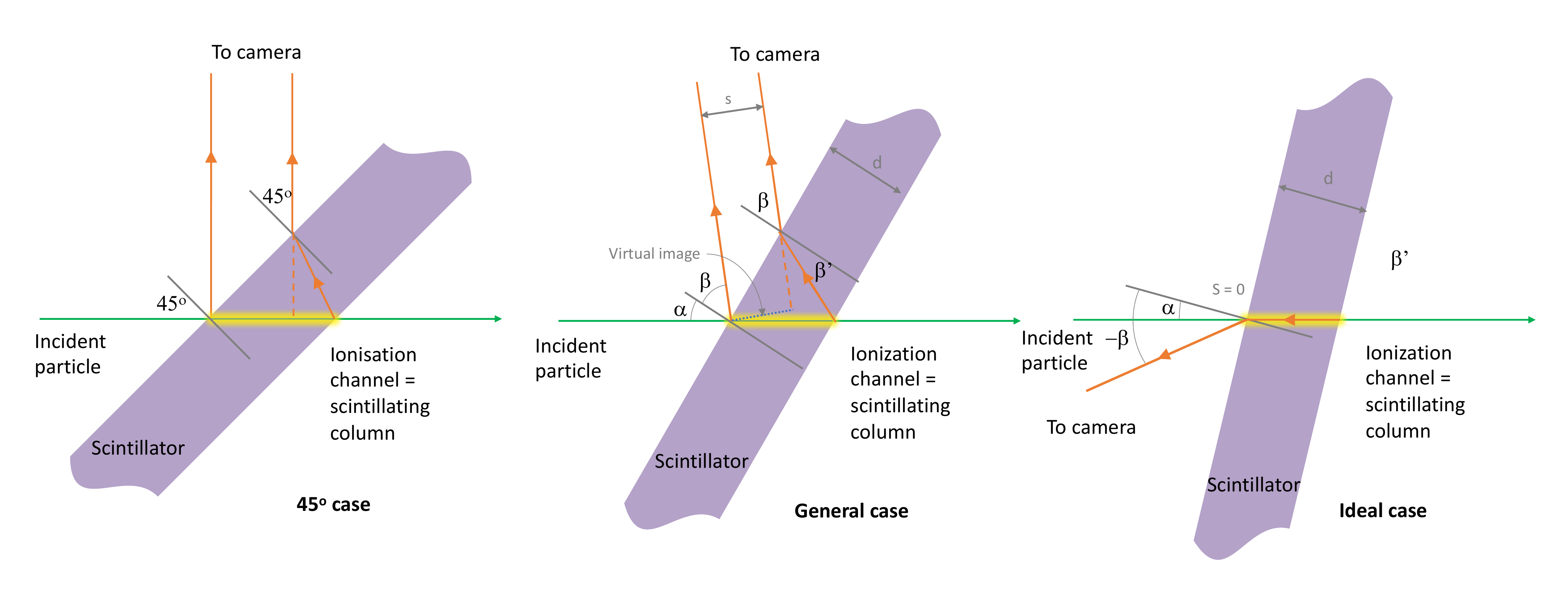}
\caption{Adjusting the tilt angle, $\alpha$, and viewing angle, $\beta$, of a thick scintillator screen can mitigate for the length of scintillating column by using refraction to improve the resolution.}
\label{fig:scintillator}
\end{center}
\end{figure}

Charged particles traversing the vacuum / screen interface may generate optical transition radiation that can be reflected towards the camera, which is is considered as a background to the scintillation light.  Each particle that crosses the scintillator creates an ionisation channel, from which light is emitted isotopically within the volume, as shown in Fig.~\ref{fig:scintillator}. Refraction of this light at the boundary affects the~virtual image size and achievable resolution. If the scintillator screen is thick compared to the desired resolution of the instrument, then an adjustment to the tilt angle with respect to the beam axis and the~viewing angle can improve the achievable resolution. As derived elsewhere~\cite{bib:Ischebeck2015}, the virtual image size, $s$, can be calculated as follows.
\begin{equation}
s =d \cos \beta \sqrt{\frac{1}{1-\frac{\sin^2 \beta}{n^2}} + \frac{1}{\cos^2 \alpha} - 2\frac{\cos \left [\arcsin (\frac{\sin\beta}{n}) + \alpha \right ]}{\sqrt{1 - \frac{\sin^2 \beta}{n^2} \cos \alpha} }}.
\end{equation}
The ideal case occurs when the tilt angle $\alpha$ and viewing angle $\beta$ are such that the virtual image size $s$ is minimised due to refraction;
\begin{equation}
\beta_{ideal} = -\arcsin ( n \sin \alpha).
\end{equation}

\subsection{Interference}
\subsubsection{Basic principles}
So far we have considered optical arrangements based on simple geometrical rays of light, however, the wave nature of light gives rise to interference effects that must at least be accounted for, or can be beautifully exploited in optical beam instrumentation. When light from a coherent, monochromatic source takes multiple optical paths to arrive at the same point in time and space, we observe interference due to the relative phase advance of each path, $\delta = \frac{2\pi}{\lambda}nD$. Consider two sinusoidal disturbances arriving at a point at time $t$, having travelled different distances, $x_1$ and $x_2$; we define the two complex spatial amplitudes, or phasors as:
 \begin{equation} 
 E_1 = a_1 \me{i\phi_1} \quad\mbox{and}\quad E_2 = a_2 \me{i\phi_2}\,,
\end{equation}
where $\phi_1 = \omega t - \delta_1$ and $\phi_2 = \omega t - \delta_2$ are the instantaneous phases.  The superposition principle states that the resulting disturbance is the sum of these complex spatial amplitudes, the phasor sum, $E = E_1 + E_2$. The measured intensity is therefore:
\begin{align}
I &= |E_1 + E_2|^2\\
 &= |E|^2  = a_1^2 + a_2^2 + 2 a_1 a_2 \cos(\delta_2 - \delta_1) \\
 &= 4a^2 \cos^2 \left ( \frac{\delta_2 -\delta_1}{2} \right),\: \mbox{for identical amplitudes }a_1 = a_2\,.
\end{align}
Note that the measured intensity depends directly on the phase difference, and hence the optical path difference (o.p.d.), which results in constructive interference for o.p.d. $= m \lambda$ and destructive interference for o.p.d. $= (m+1/2)\lambda$. \emph{Interference by division of wavefront} occurs when, for example, light passes through two infinitesimal slits and produces cosinusoidal interference fringes in the far field.

\subsubsection{Michelson Interferometer and Frequency Scanning Interferometry}
\begin{figure}[h]
\begin{center}
\includegraphics[width=0.9\textwidth]{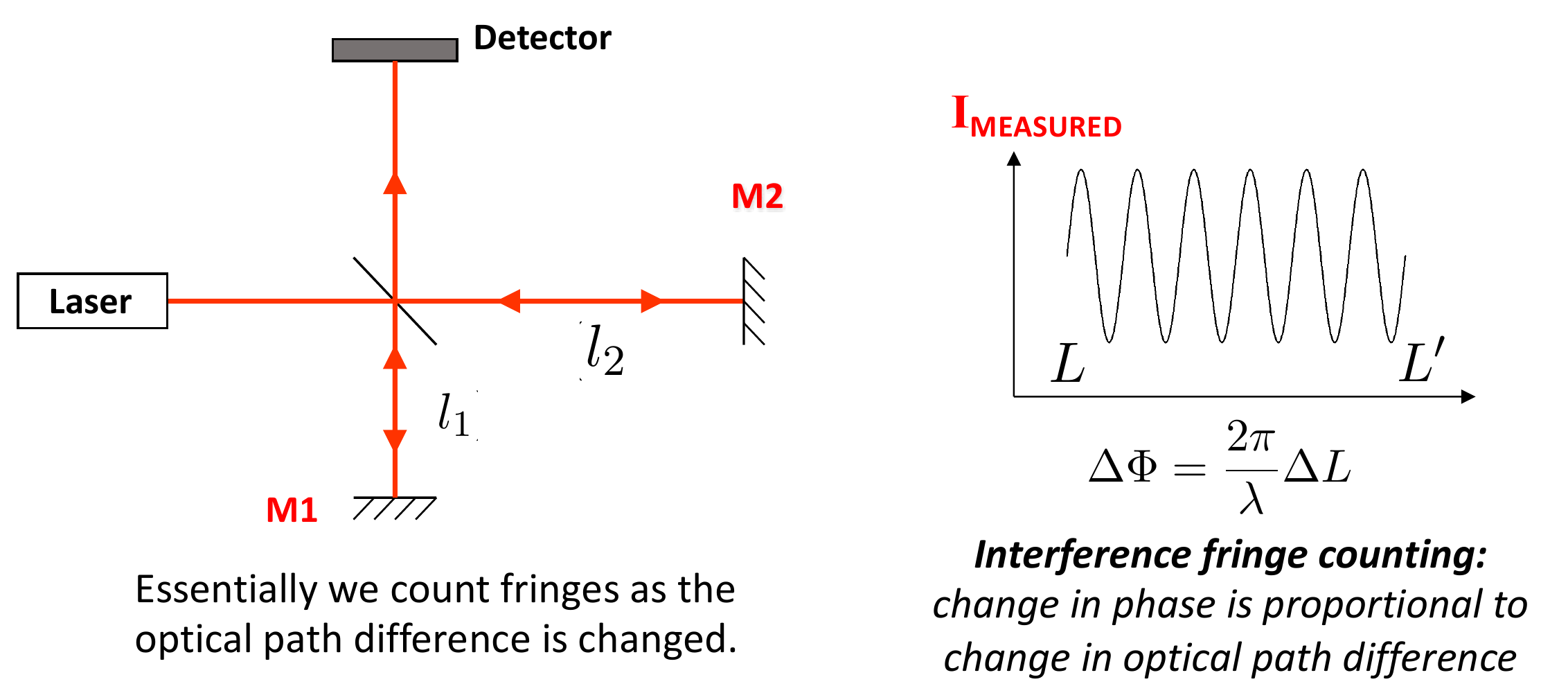}
\caption{Michelson Interferometer and fringes produced by a change in optical path difference}
\label{fig:Michelson}
\end{center}
\end{figure}

\emph{Interference by division of amplitude} occurs when light strikes a partially reflective mirror, or beam-splitter and two wavefronts are produced that are later recombined. This is the basis of the \emph{Michelson Interferometer}, shown in Fig.~\ref{fig:Michelson}, that is used widely for precise displacement measurements.
Other interferometer layouts include the \emph{Mach-Zehnder} (having one beam splitter, and a separate beam combiner), the \emph{Fabry-P\'erot etalon} (multiple reflections in an optical cavity formed between two mirrors) and \emph{Fizeau} (interference between parallel surfaces)~\cite{bib:Hecht}.

Michelson interferometry assumes that the laser frequency $\nu$ is fixed during the measurement and that only the optical path difference $L$ is changed. The phase in each arm of the interferometer is given by
\begin{equation}
\phi_1 = \frac{2\pi}{\lambda} l_1 \rm{\,\,\,and\,\,\,} \phi_2 = \frac{2\pi}{\lambda} l_2 ,
\end{equation}
where $l_1$ and $l_2$ are the round trip optical path lengths of each interferometer arm. The detected phase is therefore
\begin{equation}
\Phi = \frac{2\pi}{\lambda} (l_2 - l_1)  =  \frac{2\pi}{\lambda} L .
\end{equation}
Michelson interferometry enables precise changes in the optical path difference to be determined by continuous monitoring any change in the interferometer phase;
\begin{equation}
\Delta\Phi =  \frac{2\pi}{\lambda} \Delta L .
\end{equation}
An alternative method is to keep the optical path difference constant and vary the frequency of the laser so that the phase change is given by:
\begin{align}
\Phi &= \frac{2\pi}{c} \nu L \\
\Delta \Phi &\approx \frac{2\pi}{c} \Delta\nu L + \frac{2\pi}{c} \nu \Delta L \\
\Delta \Phi &\approx \frac{2\pi}{c} \Delta\nu L \:\: \mbox{ if $\Delta L = 0$} \,.
\end{align}
If multiple interferometers are simultaneously illuminated by the same source of tuneable light, then the~ratio of phase change is equal to the ratio of interferometer lengths:
\begin{align}
\Delta \Phi &= \frac{2\pi}{c} \Delta\nu L\,,\\
\Delta \Theta &= \frac{2\pi}{c} \Delta\nu D\\
\implies \frac{\Delta\Phi}{\Delta\Theta} &= \frac{L}{D} \,. 
\end{align}
Frequency scanning interferometry enables accurate and absolute distance measurements with respect to a reference length, even after the laser is power-cycled. FSI systems, originally developed for ATLAS~\cite{bib:Coe2004, bib:Gibson2004} and accelerator alignment, have in recent years been applied through commercial systems to a range of applications, including alignment of the HL-LHC crab-cavities~\cite{bib:Sosin2015}.

\subsection{Diffraction}
\subsubsection{Fresnel Diffraction}
\begin{figure}[h]
\begin{center}
\includegraphics[width=0.62\textwidth]{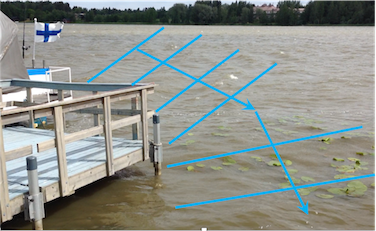}
\caption{Diffraction of water waves observed during the CAS around a boat moored on Lake Tuusula, Finland (video in lecture slides)}
\label{fig:FinnishDiffraction}
\end{center}
\end{figure}

When light meets an obstacle or aperture that is comparable in scale to the wavelength then diffraction effects become noticeable. During the CERN Accelerator School near Lake Tuusula, water waves were observed to diffract around a boat moored by the jetty outside the lecture theatre, as in Fig.~\ref{fig:FinnishDiffraction} and several other water wave analogies were explored~\cite{bib:Logiurato2012}.
In optical beam instrumentation diffraction effects can limit the resolution of the optical system, or may be exploited to directly measure the particle beam size. 

\begin{figure}[h]
\begin{center}
\includegraphics[width=0.48\textwidth]{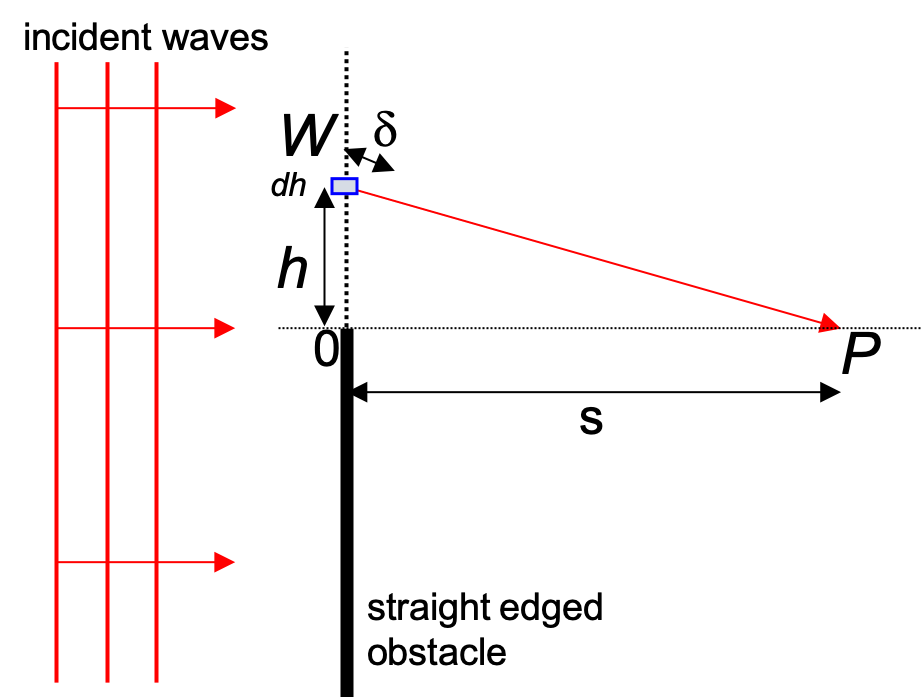}
\includegraphics[width=0.51\textwidth]{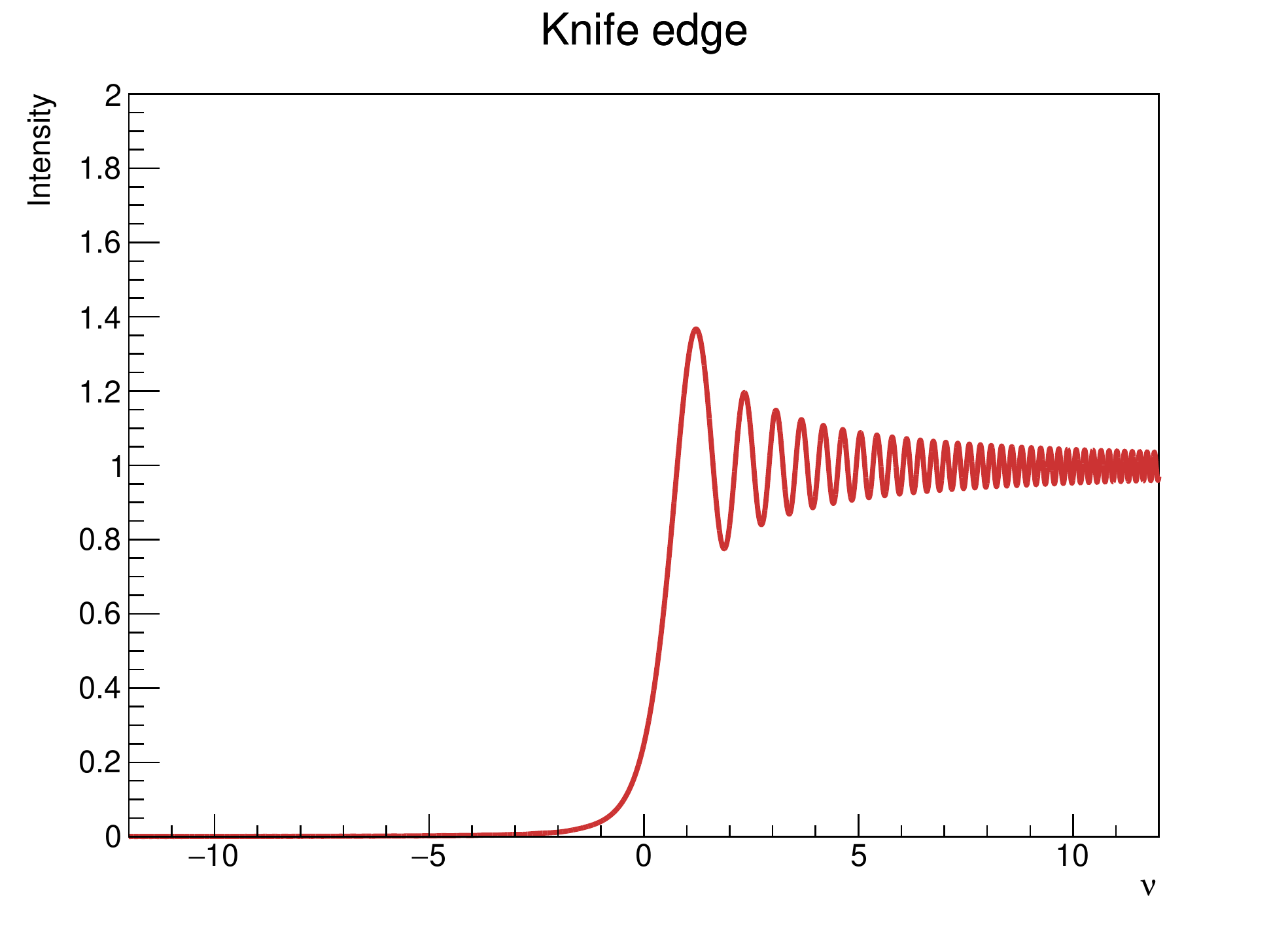}
\caption{When light waves meet a knife edge obstacle, the diffraction pattern may be calculated by integrating contributions of each unobstructed part of the wavefront. Note that in addition to the fringes, the diffracted light encroaches into the geometric shadow of the knife edge.}
\label{fig:StraightEdge}
\end{center}
\end{figure}

\begin{figure}[h]
\begin{center}
\includegraphics[width=0.55\textwidth]{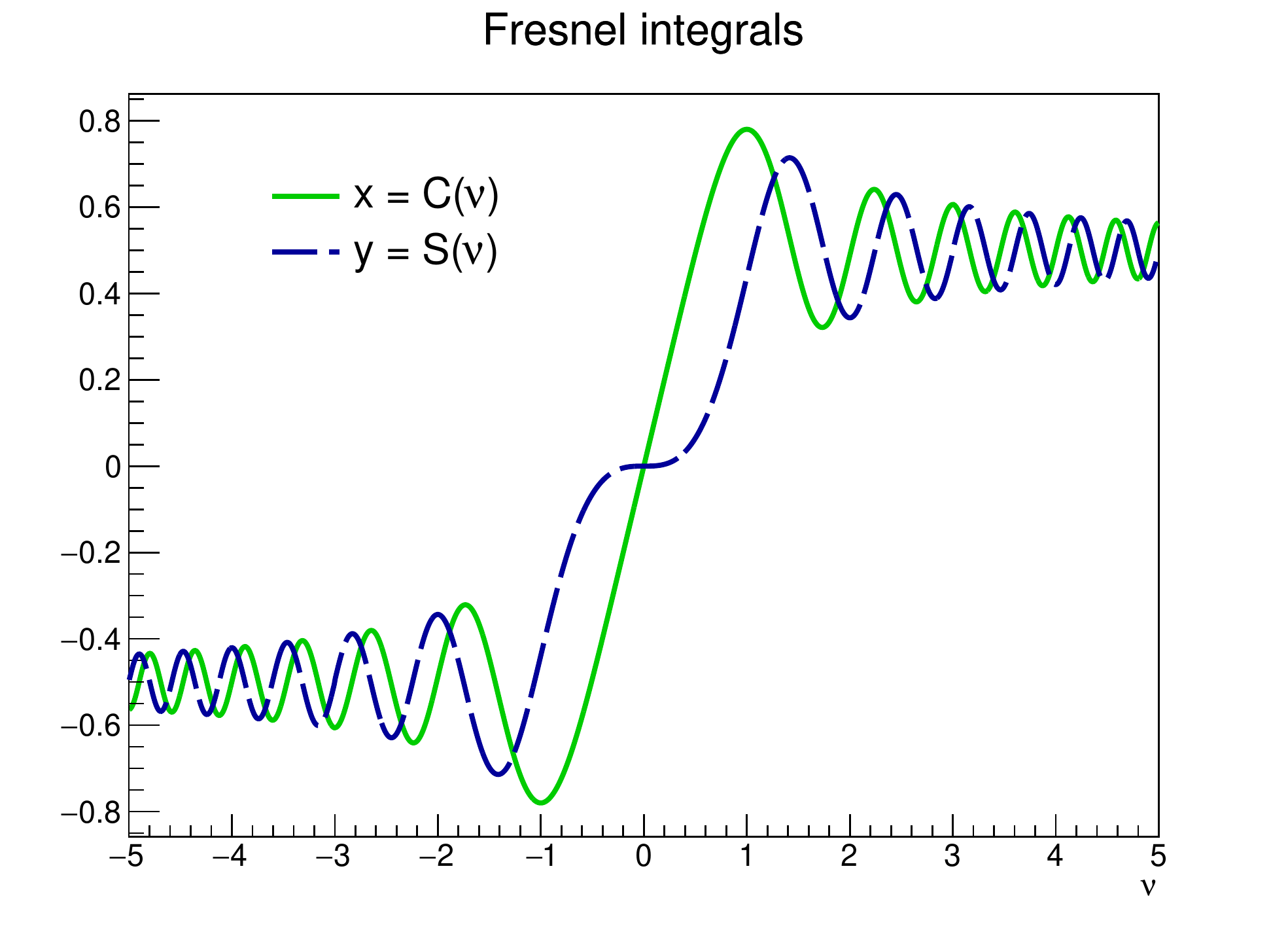}
\includegraphics[width=0.4\textwidth]{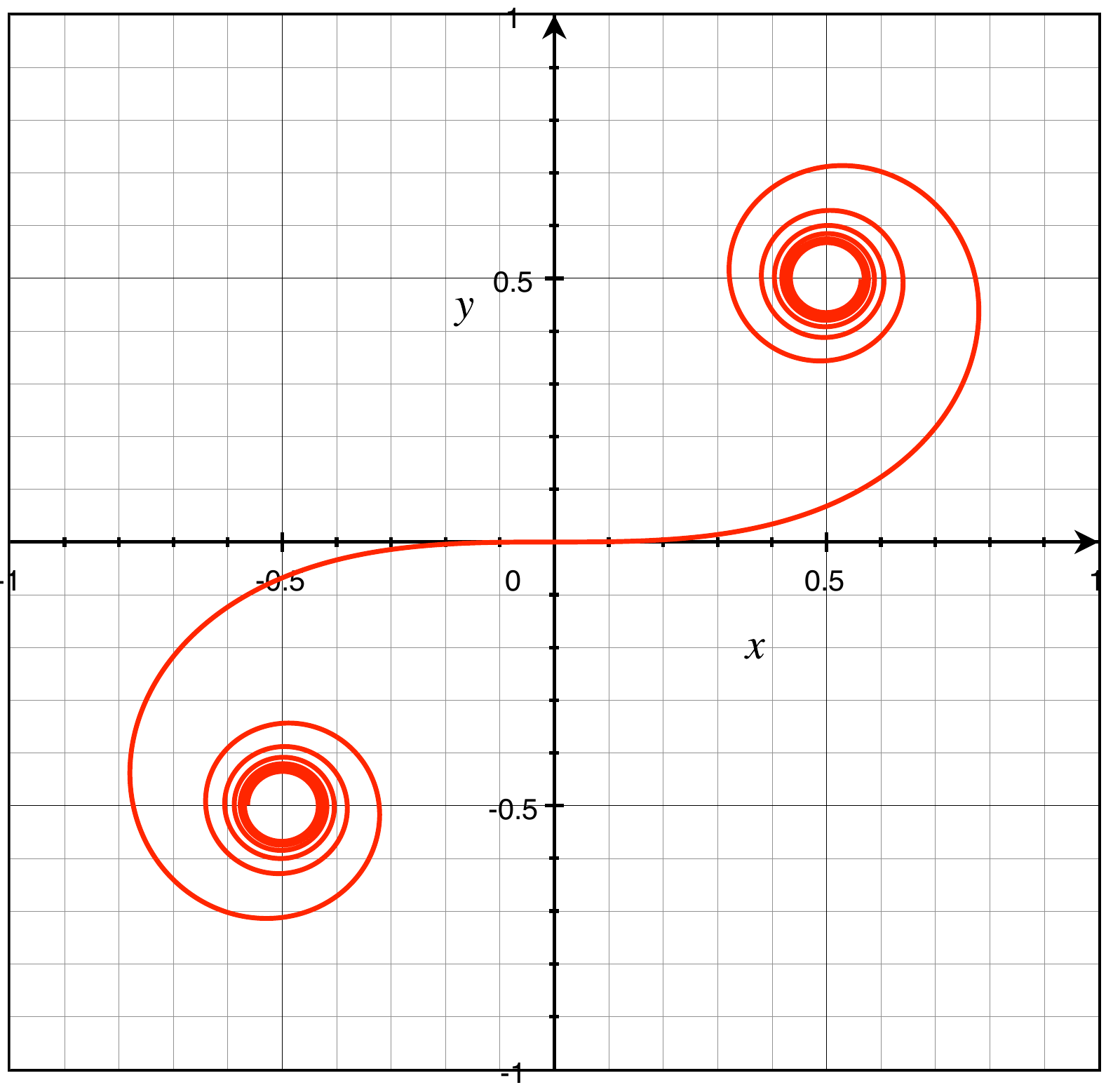}
\caption{Diffraction patterns can be calculated from the chord length between points on the Cornu spiral, with x-y coordinates defined by the Fresnel integrals of Eq.~(\ref{eqn:fresnelintegrals}).}
\label{fig:Cornu}
\end{center}
\end{figure}

Diffraction effects can be calculated by considering plane waves incident on an obstacle and integrating the contributions from the parts of the wavefront that are not obstructed, to evaluate the intensity at point P on a screen distant S from the obstacle, as shown in~Fig.~\ref{fig:StraightEdge}. 
Compared to the phase of a wave propagating from O to P, the extra phase of the wave from W is
\begin{equation}
\phi(h) = \frac{2\pi}{\lambda}\left [(s^2 + h^2)^{1/2} - s \right ] \approx \frac{\pi h^2}{\lambda s}\,,
\end{equation}
which is valid for $h^2 \ll s^2$. Note that in this Fresnel geometry, the phase is non-linear in $h$, instead of linear, as would be the case in Fraunhofer diffraction limit. We then construct a phasor, $dx+i dy = dh e^{i\phi(h)}$, due to an infinitesimal strip of height $dh$ at $h$, with components:
\begin{equation}
dx = dh \cos \left (\frac{\pi h^2}{\lambda s} \right ) \quad \mbox{and} \quad dy = dh \sin \left (\frac{\pi h^2}{\lambda s} \right )
\end{equation}
The phasor traces out the Cornu spiral on the Argand diagram, shown in Fig.~\ref{fig:Cornu}, with coordinates that are given by the Fresnel integrals:
\begin{equation}\label{eqn:fresnelintegrals}
x = \int^\nu_0 \cos \frac{\pi \nu^{\prime 2}}{2} d \nu^\prime \quad \mbox{and} \quad y = \int^\nu_0 \sin \frac{\pi \nu^{\prime 2}}{2} d \nu^\prime \,,
\end{equation}
where $\nu = h \left ( \frac{2}{\lambda s} \right ) ^{1/2}$ is a dimensionless variable that represents the distance along the spiral. The~amplitude of the diffraction pattern formed on the screen beyond the obstacle is determined by the length of the chord between two points on the Cornu spiral. The light intensity is found from the normalised square of the chord length.

\begin{figure}[h!]
\begin{center}
\includegraphics[width=0.32\textwidth]{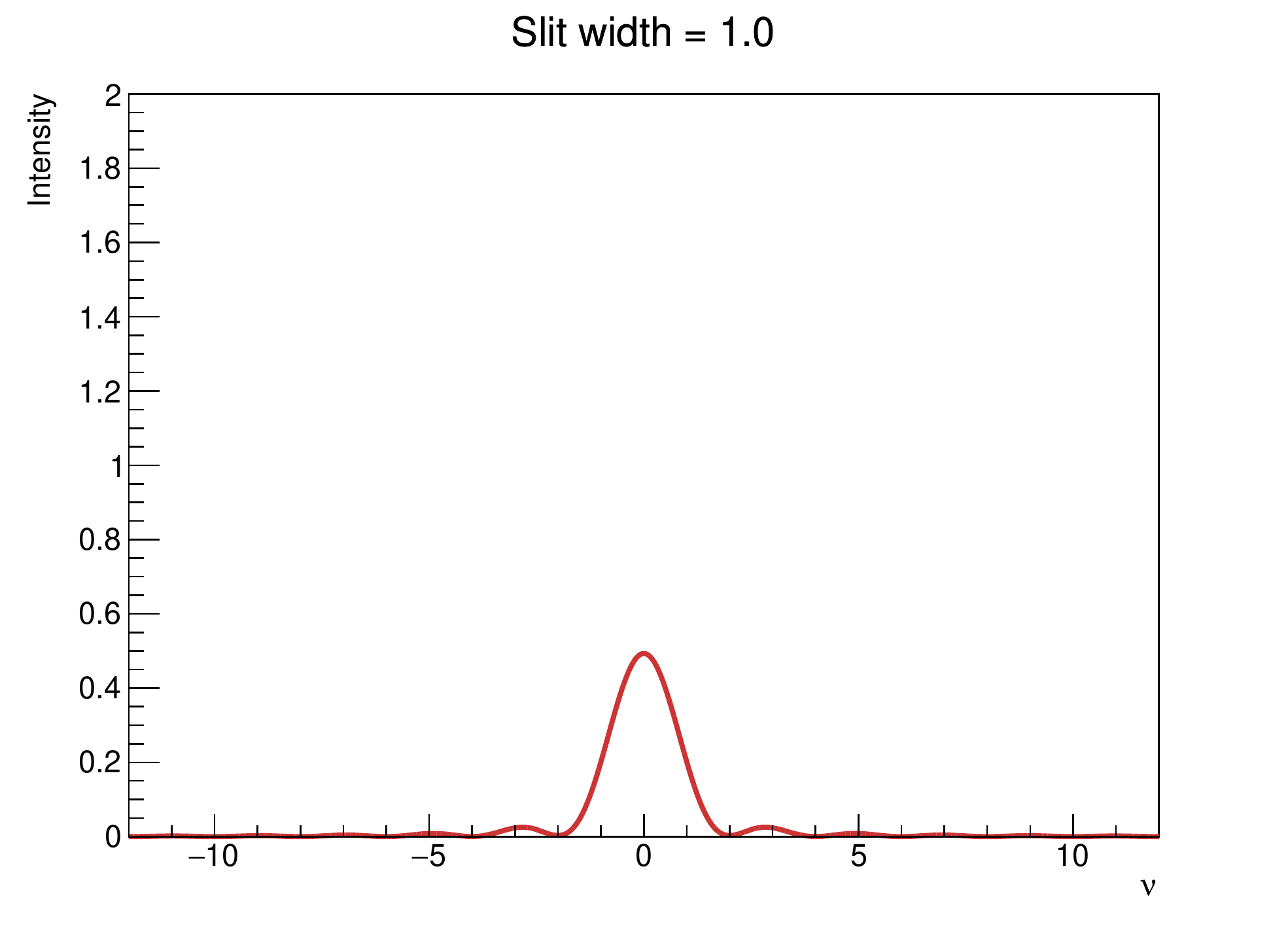}
\includegraphics[width=0.32\textwidth]{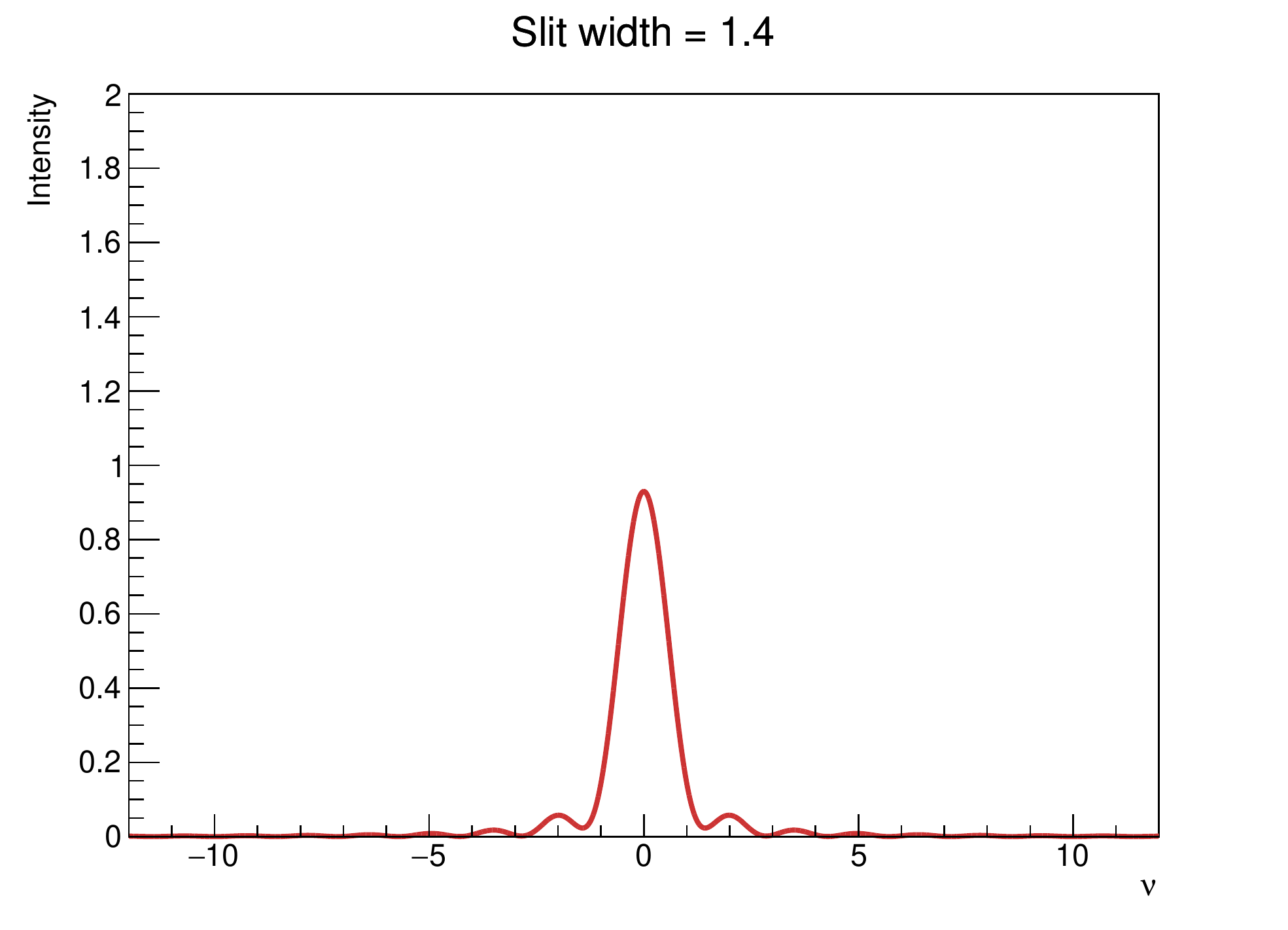}
\includegraphics[width=0.32\textwidth]{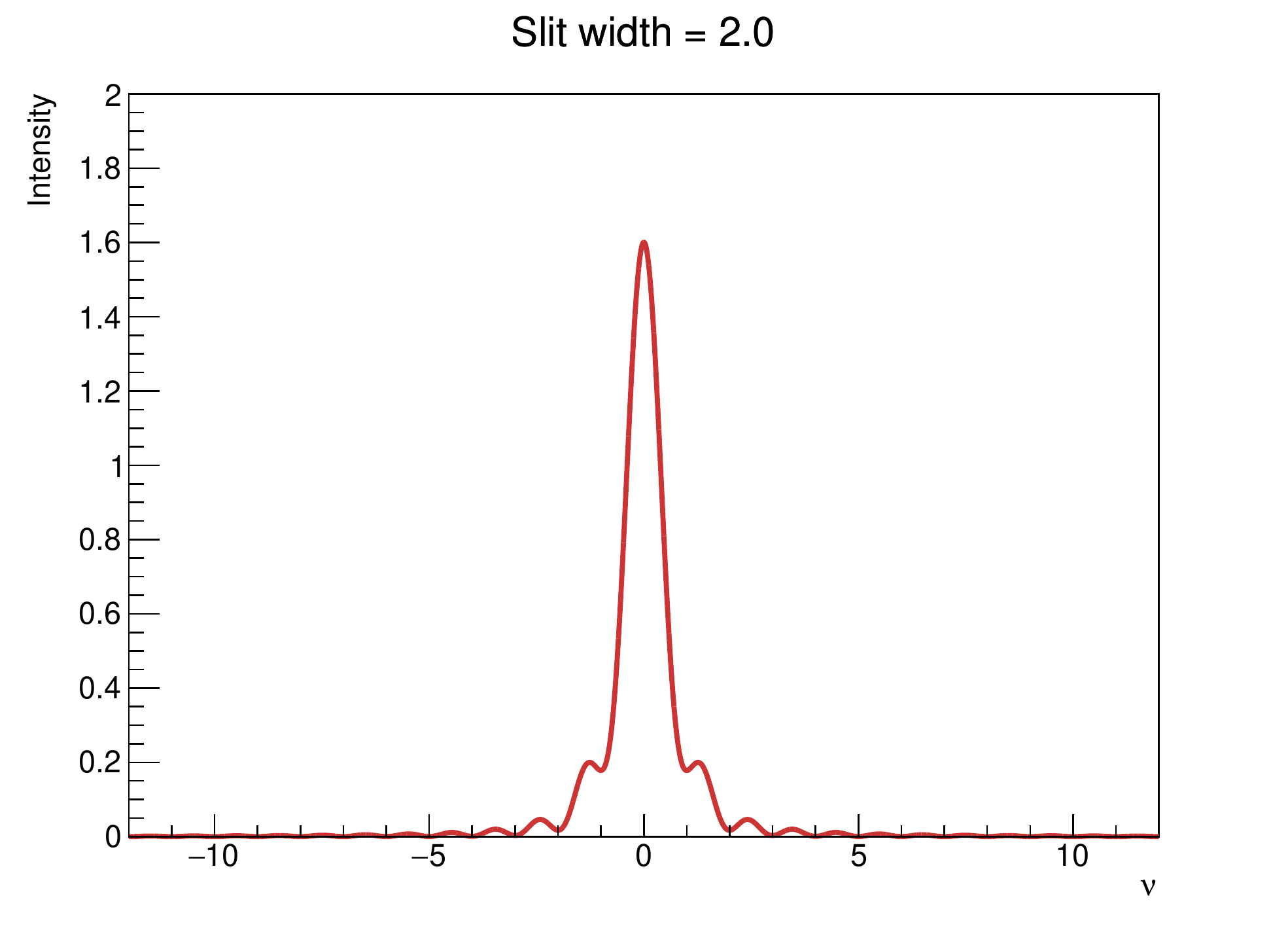}
\hrule
\includegraphics[width=0.32\textwidth]{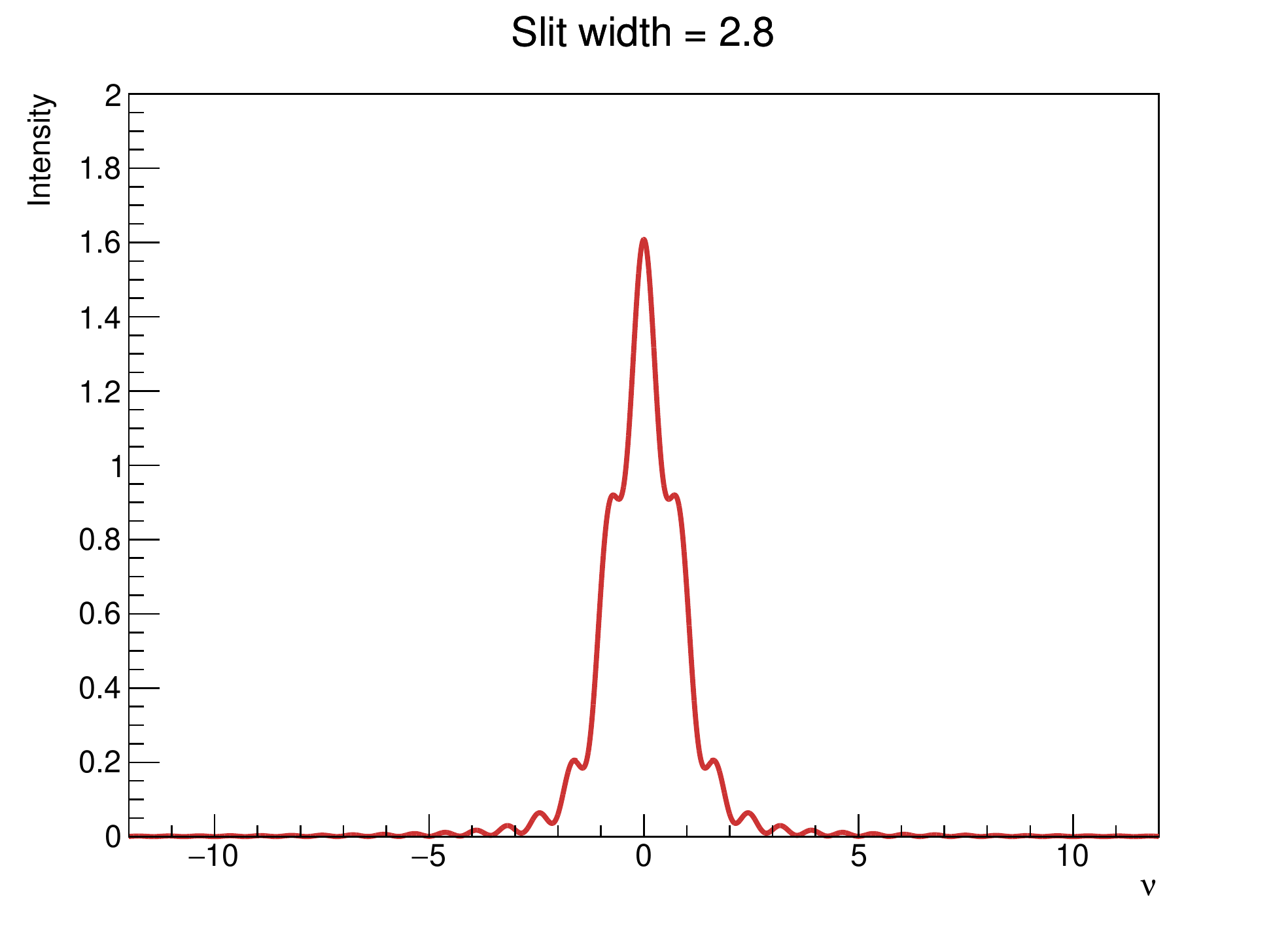}
\includegraphics[width=0.32\textwidth]{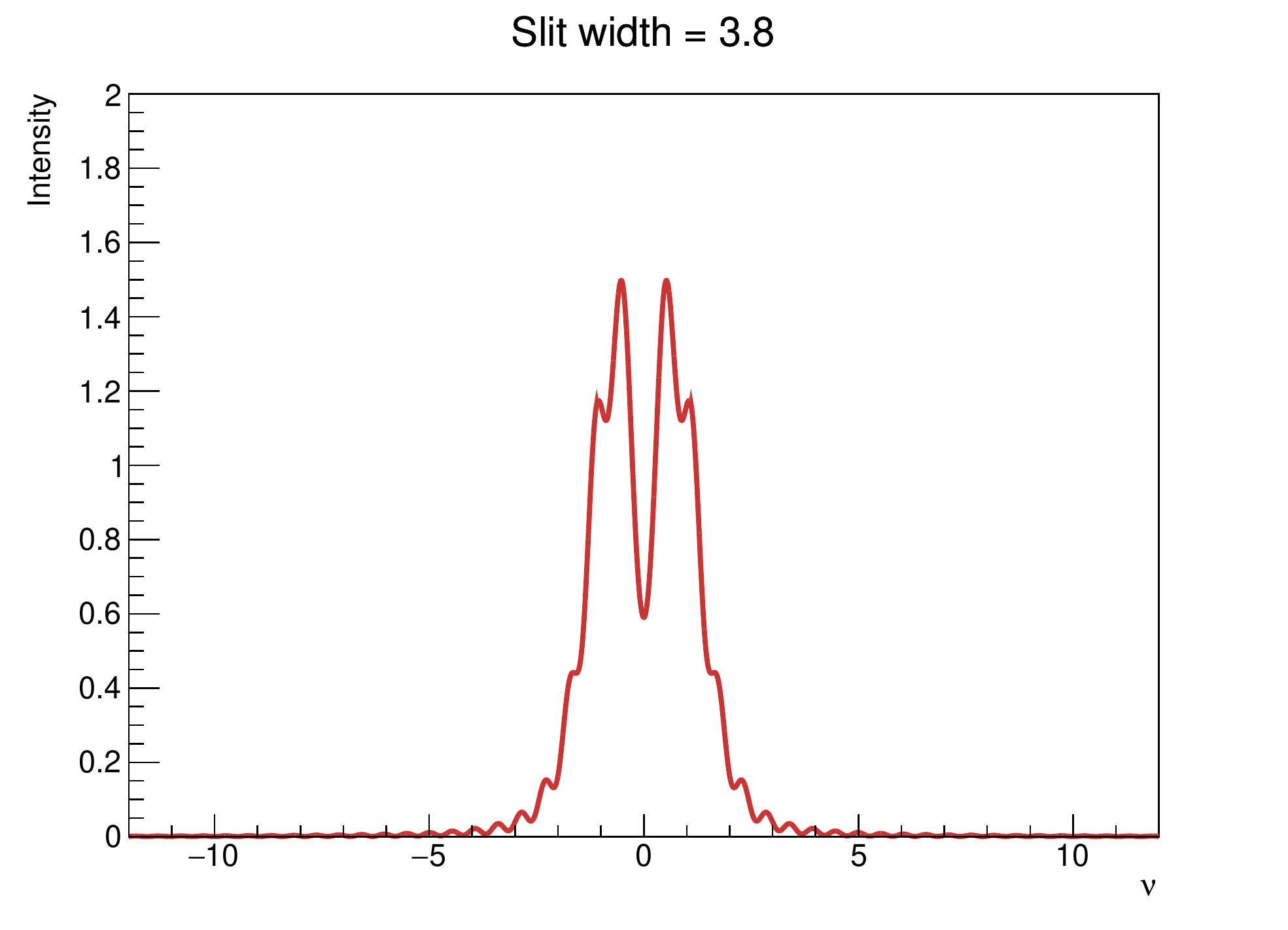}
\includegraphics[width=0.32\textwidth]{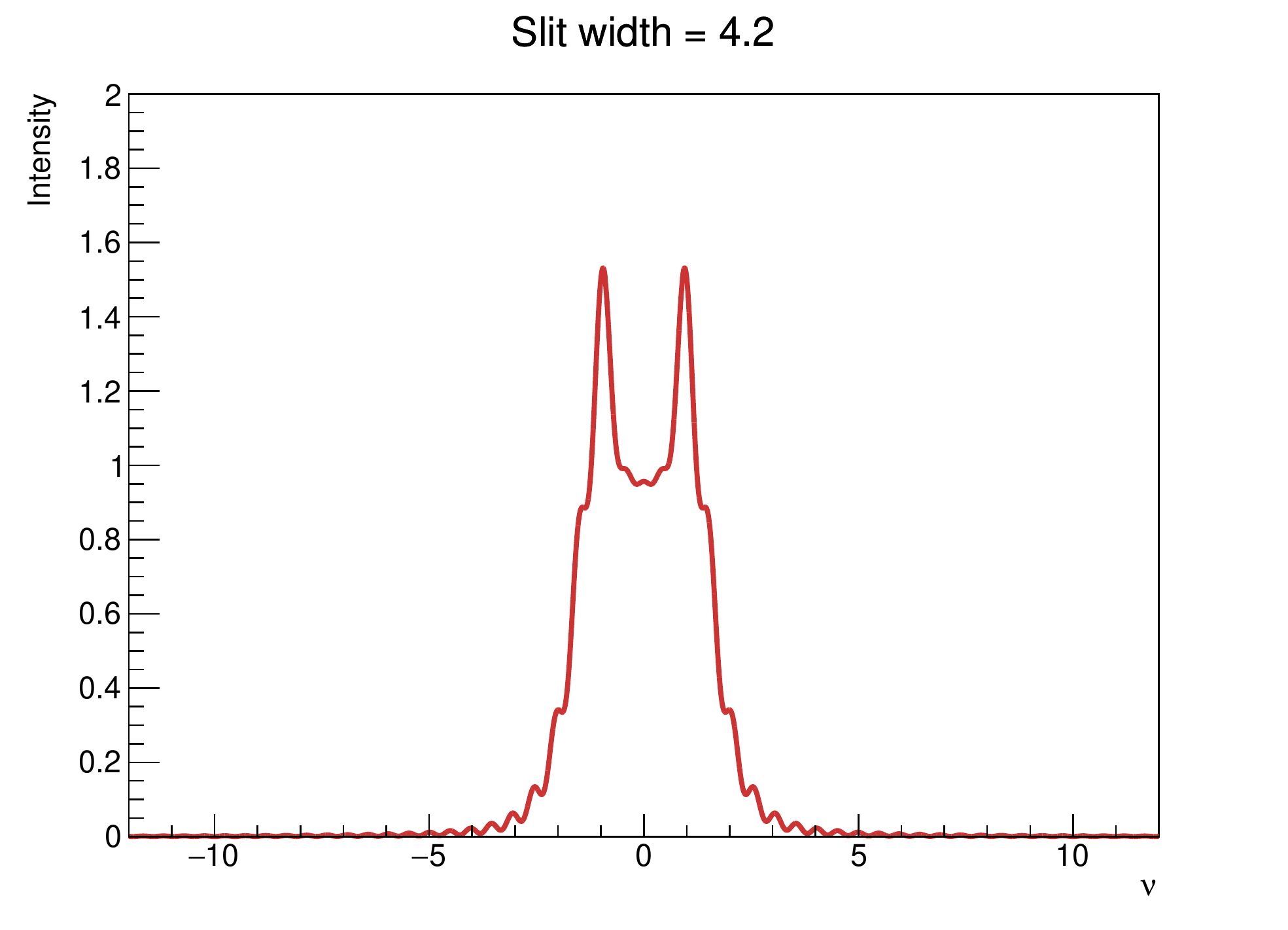}
\hrule
\includegraphics[width=0.32\textwidth]{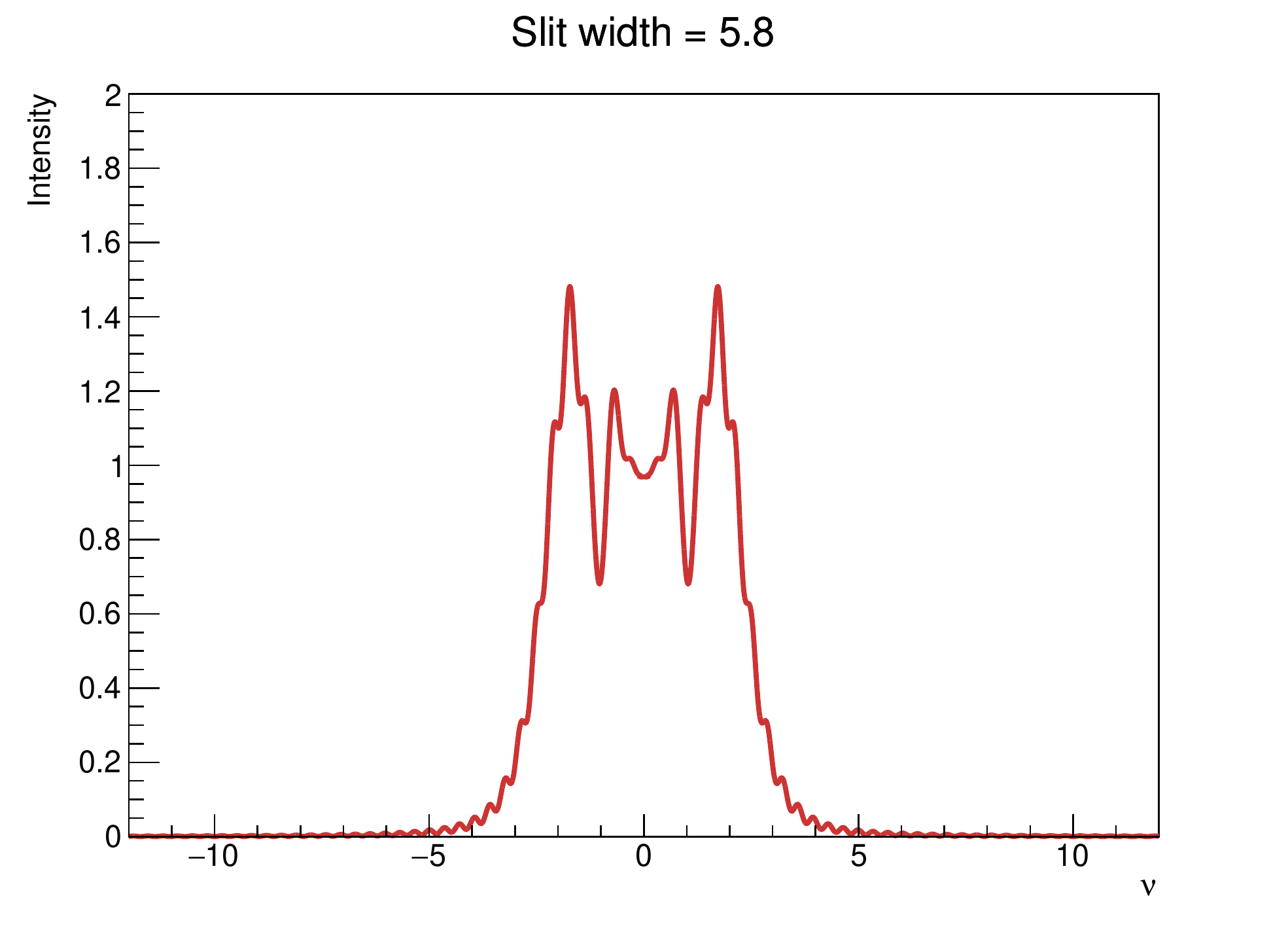}
\includegraphics[width=0.32\textwidth]{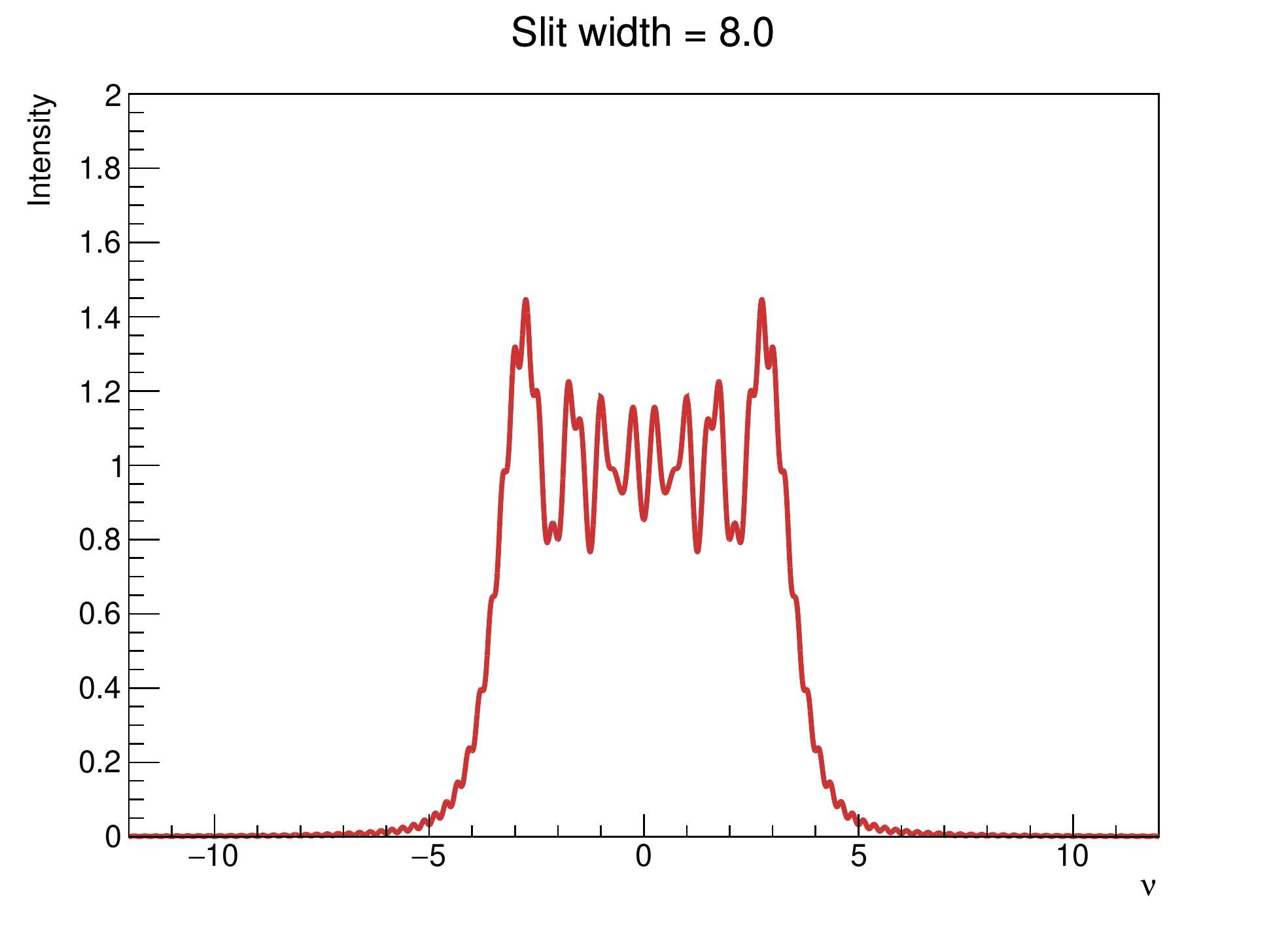}
\includegraphics[width=0.32\textwidth]{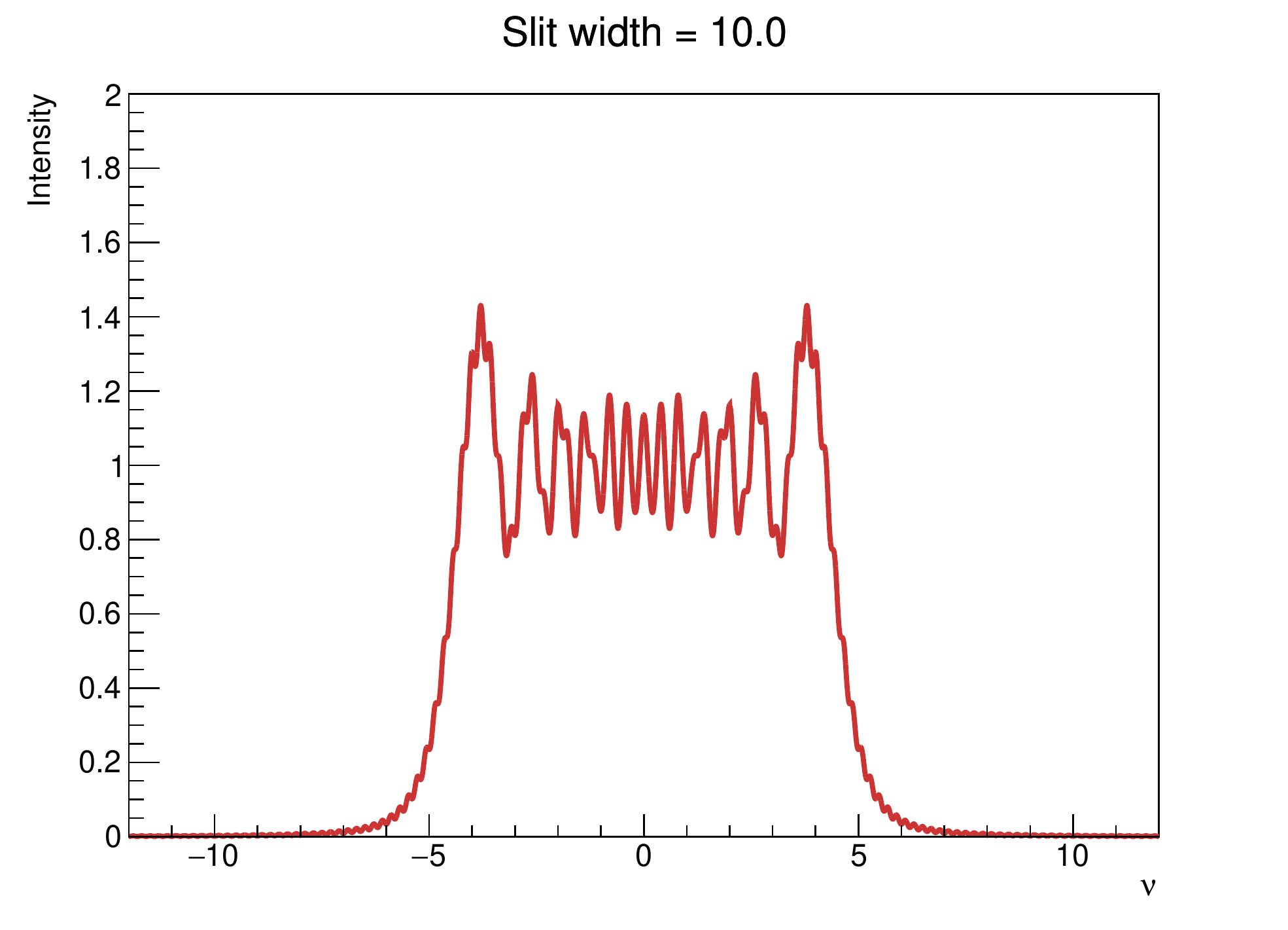}
\hrule
\includegraphics[width=0.32\textwidth]{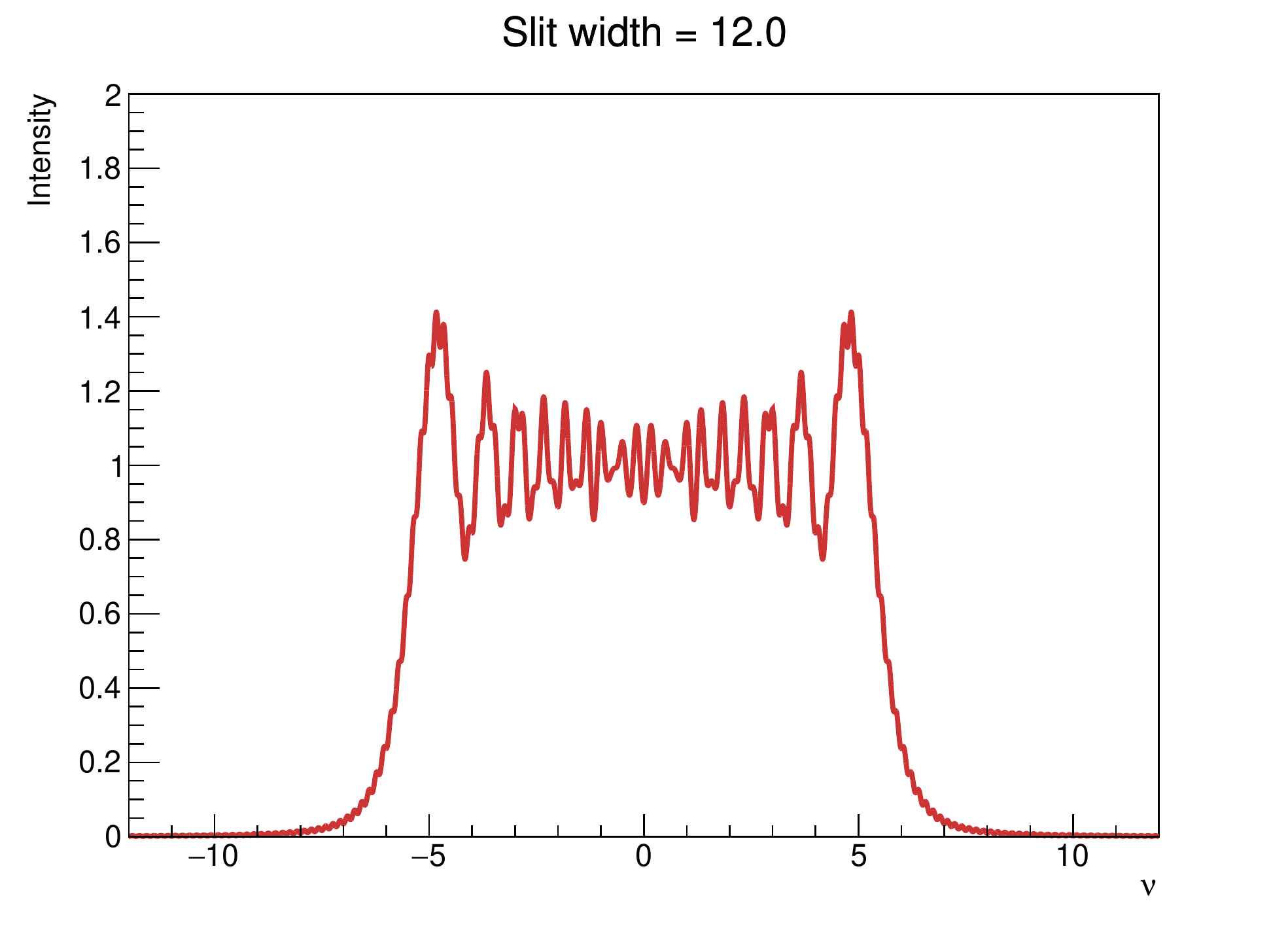}
\includegraphics[width=0.32\textwidth]{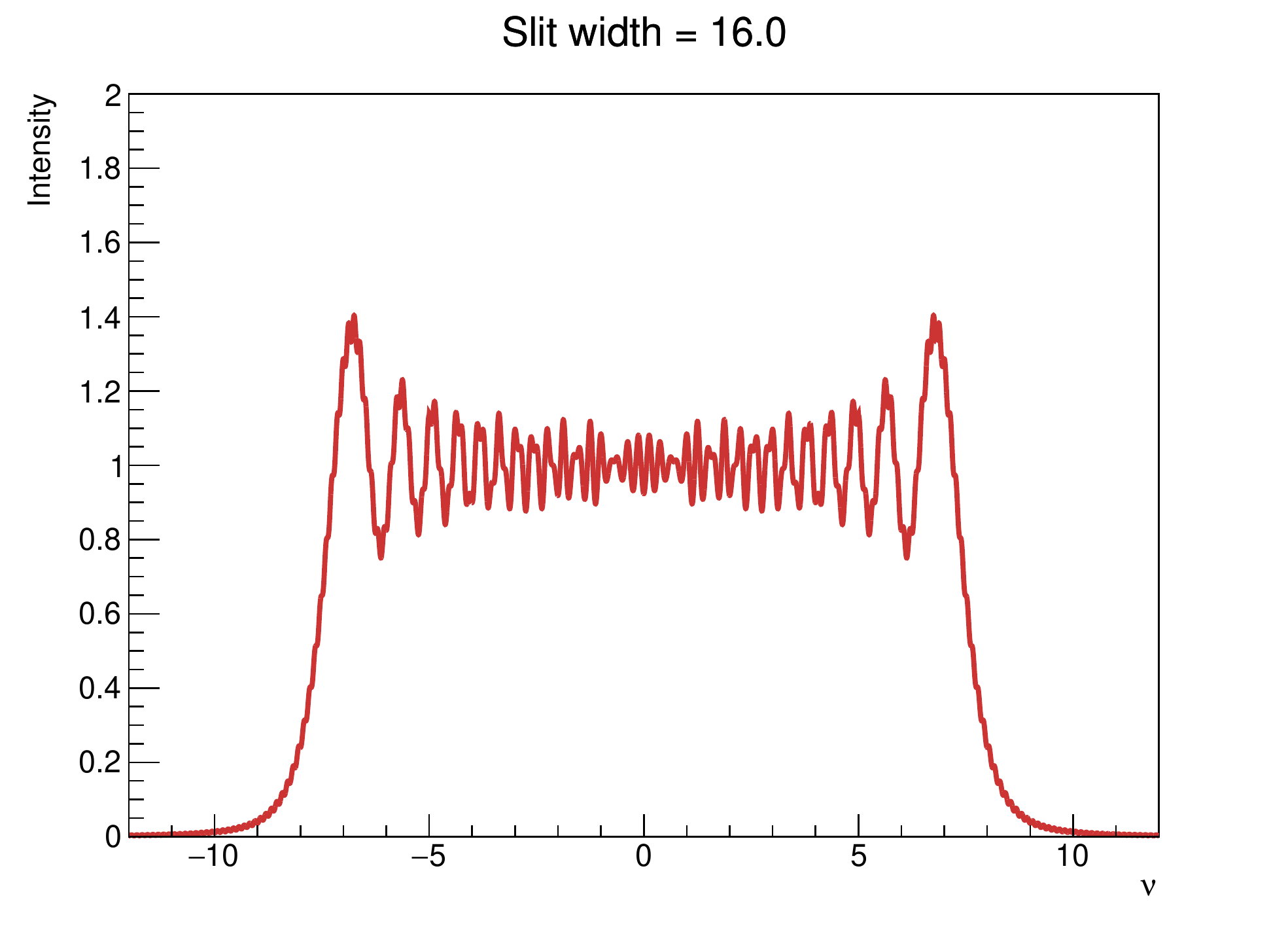}
\includegraphics[width=0.32\textwidth]{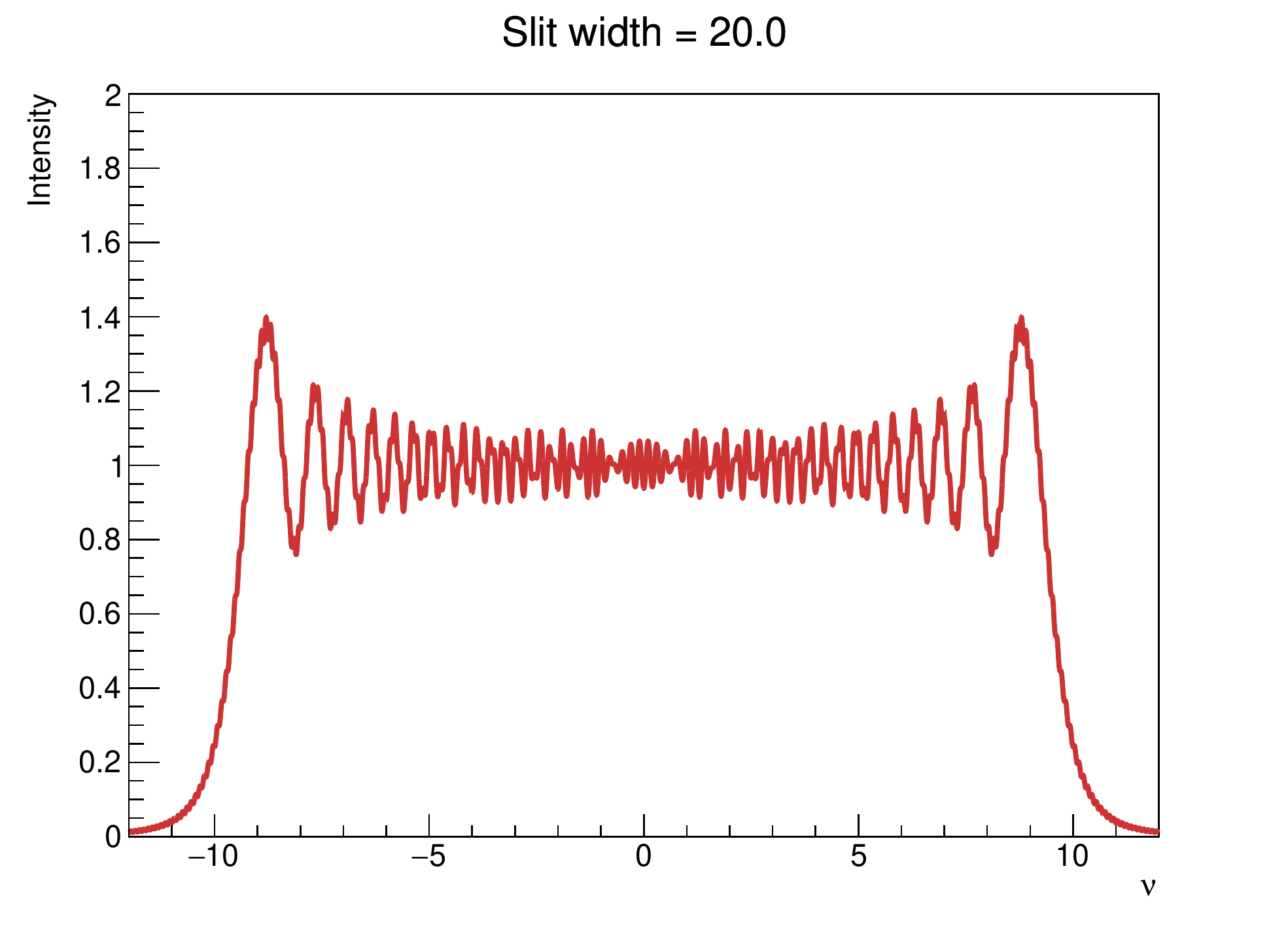}
\caption{The transition from Fraunhofer to Fresnel diffraction intensity patterns as a single slit aperture is widened}
\label{fig:FresnelFraunhofer}
\end{center}
\end{figure}

A graphical example is given in Fig.~\ref{fig:FresnelFraunhofer}, which shows the intensity pattern beyond a single slit aperture calculated from the Cornu spiral, for a variety of different slit widths. As the slit is widened, the~pattern transforms from the well known Fraunhofer single slit ($sinc^2 x$) pattern, through various Fresnel patterns with dark central minima, to the Fresnel patterns effectively formed by two well separated opposing straight knife-edges. The diffraction regime depends on the relative size of the aperture, wavelength and screen distance, according to the Fresnel number, $F = \frac{a^2}{S \lambda} \ll 1$ for Fraunhofer diffraction.

\subsubsection{Fraunhofer diffraction and convolution theorem}
In the Fraunhofer limit, the general method to calculate the far field diffraction pattern is to take the~Fourier Transform (FT) of the transmission function of the diffracting aperture. 
\begin{equation}
I(\theta_x) = | E_{res} (\theta_x) | ^2 = \left | \int_s A(x_s) e^{-i k x_s \sin \theta_x } dx_s \right |^2
\end{equation}
Calculated diffraction patterns resulting from several common apertures are presented in Table~\ref{tab:DiffractionTable}.
\begin{table}[h!]
\caption{Solutions to common one- and two- dimensional diffracting aperture functions}
\label{tab:DiffractionTable}
\centering\small
\begin{tabular}{cc}
\hline\hline
\textrm{\emph{\bfseries Single slit aperture}} &   \emph{\bfseries 1D Fraunhofer pattern}\\\hline
 & \multirow{3}{*}{\includegraphics[width=0.33\textwidth]{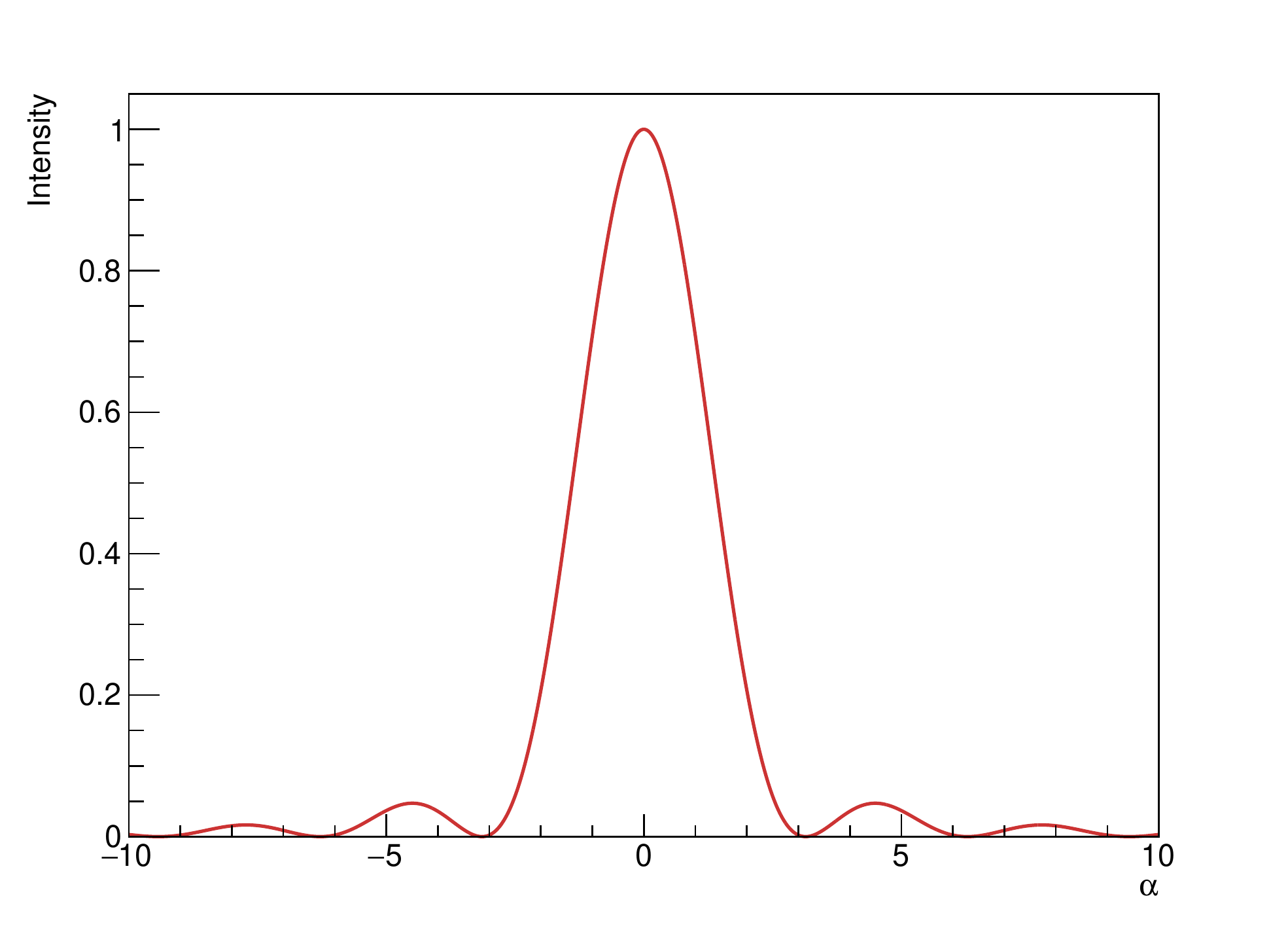}} \\
$I(\theta_x) = | E_{res} (\theta_x) | ^2 = \left | \int_{-a/2}^{a/2} A_0 e^{-i k x_s \sin \theta_x } dx_s \right |^2$ \\
for slit width $a$.\\
\\
Solution: $I(\theta_x) = A_0^2 \frac{\sin^2{\alpha}}{\alpha^2}$,\\
\\
where $\alpha = \frac{\pi}{\lambda} a \sin{\theta_x}$.\\
\\
\hline
\textrm{\emph{\bfseries Double slit aperture}} &   \emph{\bfseries 1D Fraunhofer pattern}\\\hline
 & \multirow{3}{*}{\includegraphics[width=0.33\textwidth]{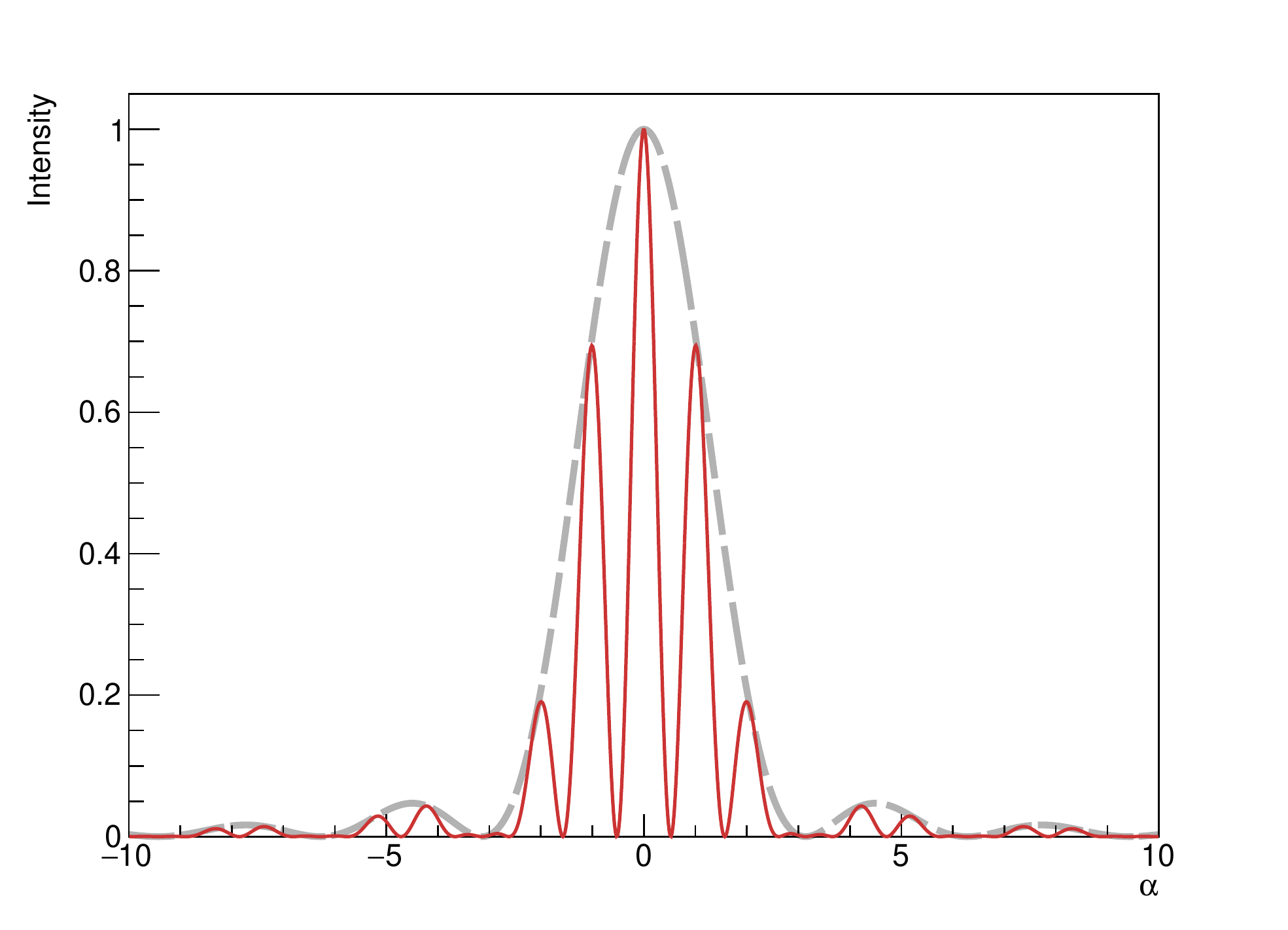}} \\
$I(\theta_x) = \left | \int_{-d/2 - a/2}^{-d/2+a/2} A_0 e^{-i k x_s \sin \theta_x } dx_s + \int_{d/2-a/2}^{d/2+a/2} A_0 e^{-i k x_s \sin \theta_x } dx_s\right |^2$ \\
for slit width $a$ and slit spacing $d$.\\
\\
Solution: $I(\theta_x) = A_0^2 \frac{\sin^2{\alpha}}{\alpha^2} \cos^2\frac{\delta}{2}$,\\
\\
where $\alpha = \frac{\pi}{\lambda} a \sin{\theta_x}$ and $\delta = \frac{2\pi}{\lambda} d \sin{\theta_x}$.\\
\\
\hline
\textrm{\emph{\bfseries N-slit grating aperture}} &   \emph{\bfseries 1D Fraunhofer pattern}\\\hline
 & \multirow{3}{*}{\includegraphics[width=0.33\textwidth]{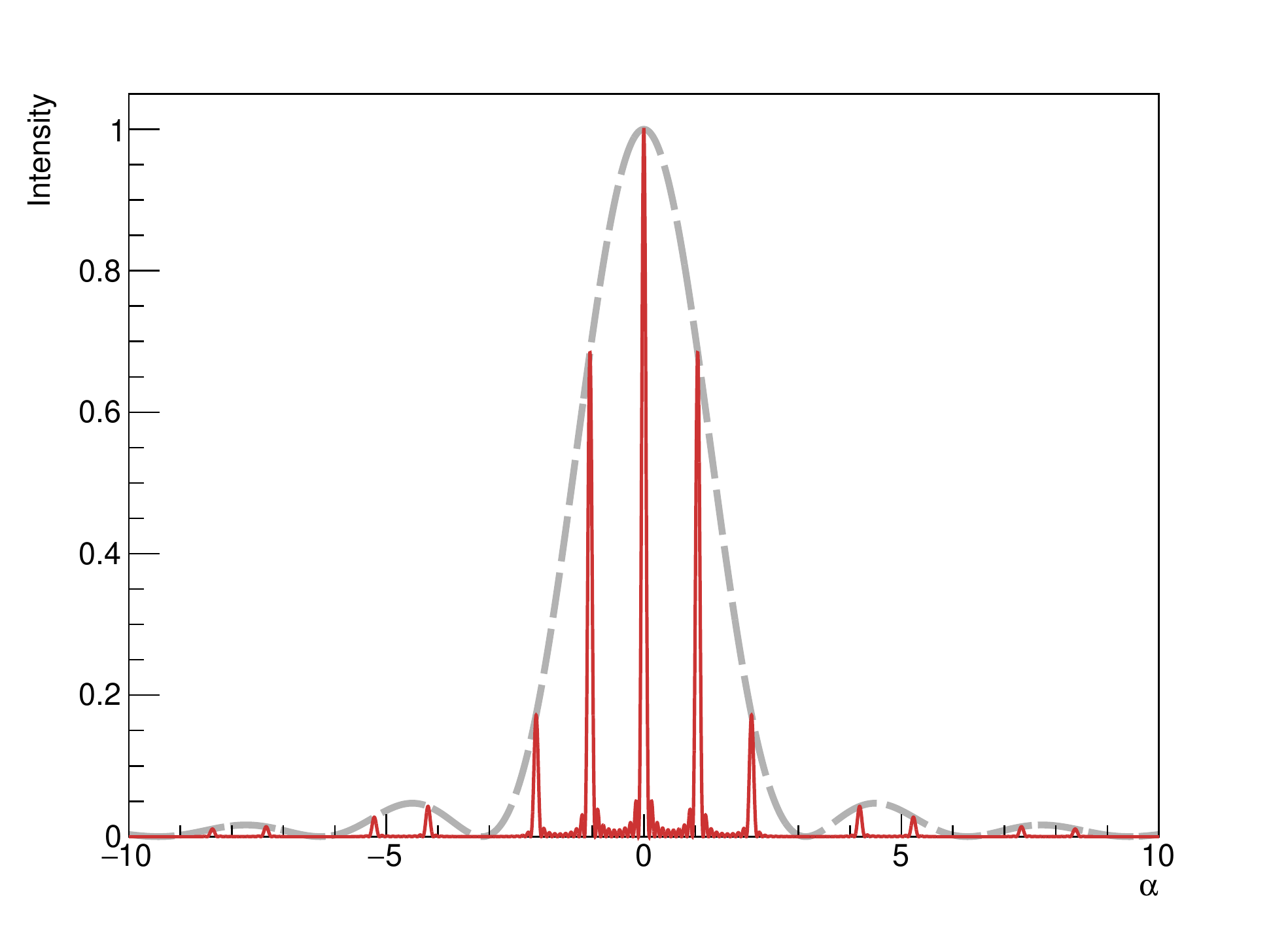}} \\
$I(\theta_x) =  | \int_{-a/2}^{a/2} A_0 e^{-i k x_s \sin \theta_x } dx_s + \int_{d-a/2}^{d+a/2} A_0 e^{-i k x_s \sin \theta_x } dx_s +...$\\
$... + \int_{(N-1)d-a/2}^{(N-1)d+a/2} A_0 e^{-i k x_s \sin \theta_x } dx_s |^2$ \\
for N slits of width $a$ and slit spacing $d$.\\
\\
Solution: $I(\theta_x) = A_0^2 \frac{\sin^2{\alpha}}{\alpha^2} \frac{\sin^2{N \beta}}{\beta^2}$,\\
where $\alpha = \frac{\pi}{\lambda} a \sin{\theta_x}$ and $\beta = \frac{\delta}{2} = \frac{\pi}{\lambda} d \sin{\theta_x}$.\\
\\
\hline
\textrm{\emph{\bfseries Rectangular slit aperture}} &   \emph{\bfseries 2D Fraunhofer pattern}\\\hline
 & \multirow{3}{*}{\includegraphics[width=0.30\textwidth]{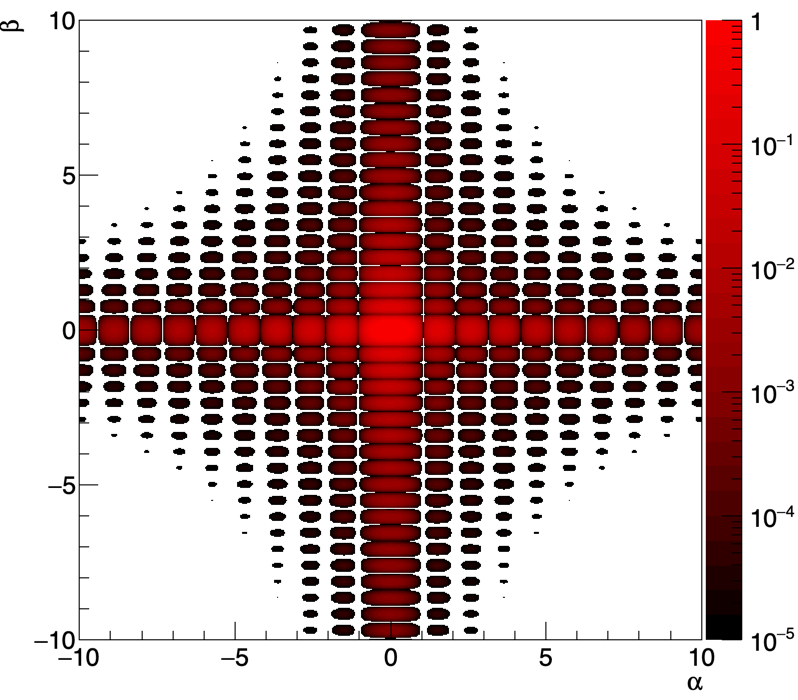}} \\
$I(\theta_x, theta_y) = \left | \int_{-a/2}^{a/2} \int_{-b/2}^{b/2} A_0 e^{-i k x_s \sin \theta_x } e^{-i k y_s \sin \theta_y } dx_s dy_s \right |^2$ \\
for slit width $a$ and slit height $b$.\\
\\
Solution: $I(\theta_x, \theta_x) = A_0^2 \frac{\sin^2{\alpha}}{\alpha^2} \frac{\sin^2{\beta}}{\beta^2}  $,\\
\\
where $\alpha = \frac{\pi}{\lambda} a \sin{\theta_x}$ and $\beta = \frac{\pi}{\lambda} b \sin{\theta_y}$.\\
In the example plot, $a < b$.\\
\\
\hline
\textrm{\emph{\bfseries Circular aperture}} &   \emph{\bfseries 2D Fraunhofer pattern}\\\hline
 & \multirow{3}{*}{\includegraphics[width=0.30\textwidth]{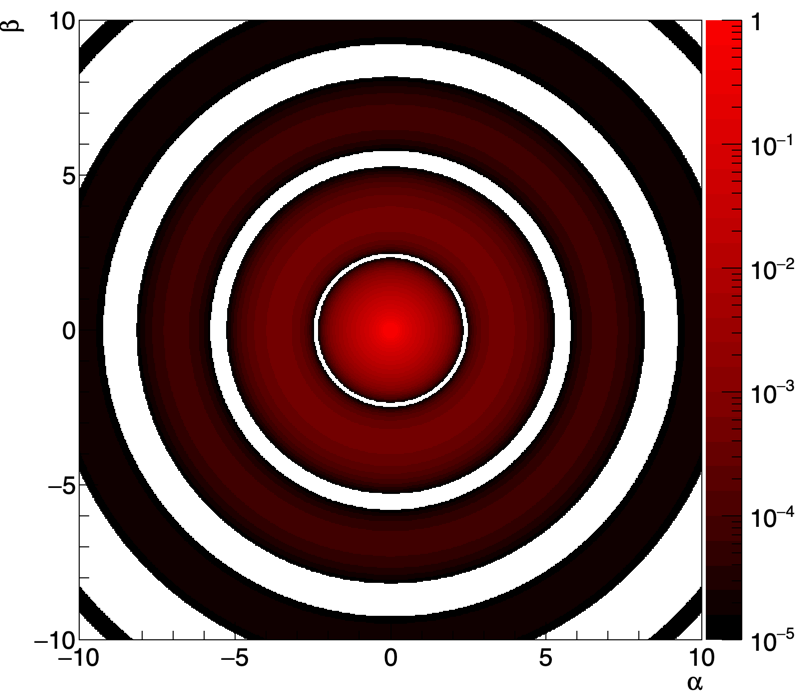}} \\
$I = \left | C e^{i k r_0} \int_{-R}^{+R}  e^{i k y \sin \theta} 2\sqrt{R^2 - y^2} dy \right |^2$ \\
for a circular aperture of diameter D =  2R.\\
\\
Solution: $I = (2CR^2)^2 \left [ \pi \frac{J_1 (\rho)}{\rho} \right ]^2 = I_0 \left [ 2 \frac{J_1 (\rho)}{\rho} \right ]^2 $,\\
\\
where $J_1$ is the first order Bessel function of the first kind and $\rho = kR \sin \theta$.\\
Note that the first dark ring occurs when  $\rho = k R \sin \theta =  3.832$,\\
or $\sin \theta =  1.22 \frac{\lambda}{D}$, which is the Rayleigh criterion. \\
\hline\hline
\end{tabular}
\end{table}
Such solutions may be extended in two dimensions by a double integral over both coordinates, for example for rectangular and circular apertures. Familiarisation with common FT pairs can be useful to identify and potentially mitigate each source of diffraction in optical setups.

\begin{figure}[h!]
\begin{center}
\includegraphics[width=0.35\textwidth]{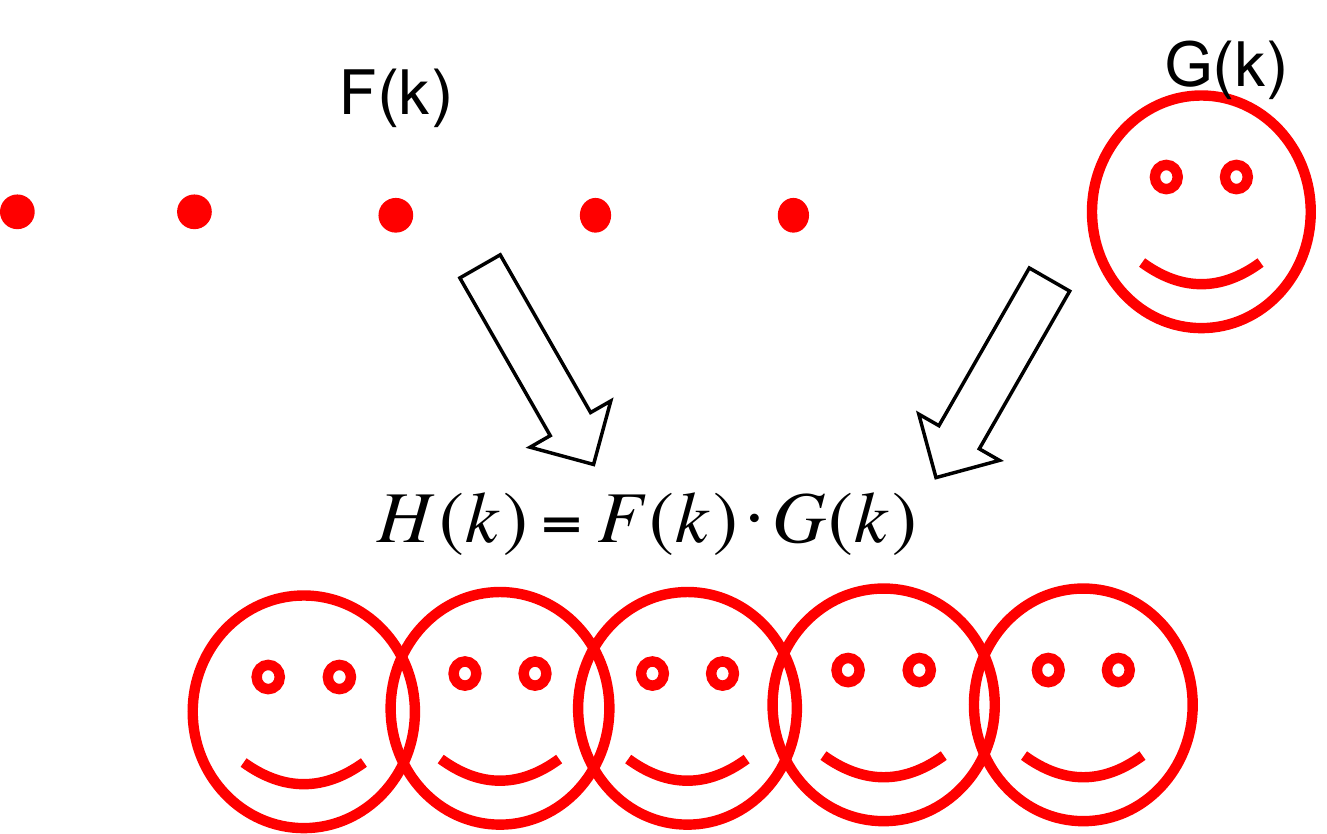}
\caption{A reminder of the lecture demonstration of convolution.}
\label{fig:convolution}
\end{center}
\end{figure}

Identification is aided by the convolution theorem, for which the convolution function is defined as
\begin{equation}
h(x) = f(x) \circledast g(x) = \int_{-\infty}^{\infty} f(x^\prime) g(x^\prime - x) dx^\prime \label{eqn:convolution}
\end{equation}
If the Fourier Transforms of $f(x)$, $g(x)$ and $h(x)$ are $F(k)$, $G(k)$, and $H(k)$ respectively, then the~convolution theorem states that
\begin{equation}
H(k) = F(k) \cdot G(k),
\end{equation}
Comparing to Eq.~(\ref{eqn:convolution}), note that the FT of a convolution of $f$ and $g$ is the product of the FTs of $f$ and $g$. In the lecture it was demonstrated that a laser beam that passes separately through an N-slit grating to give an array of points, or through a patterned grating give a smiley face, would produce an array of multiple faces, when the beam passed through both gratings, as sketched in Fig.~\ref{fig:convolution}.

\subsubsection{Mitigation of diffraction effects}
In addition to the observation geometry, the resolution at the image plane will be influenced by diffraction at any restrictive apertures, around obstructions (dust), or aberrations due to lens imperfections or refractive index variations in the optical system. The point spread function (PSF) is the optical response of the system to a single point of light at the object plane. In the case of a bunch profile measurement for example, it is the image resulting from the passage of one charged particle through the scintillator. The~PSF is independent of position in the image plane (shift theorem), so a deconvolution can be applied to enhance the resolution of the image. Deconvolution is the process of filtering a signal to compensate for an undesired convolution: in this case the convolution of the bunch profile signal with the PSF. The~goal of deconvolution is to recreate the signal as it existed before the convolution took place. Once the PSF has been modelled or directly measured, the deconvolution of the image can be achieved through digital signal processing.

\begin{figure}[h!]
\begin{center}
\includegraphics[width=0.75\textwidth]{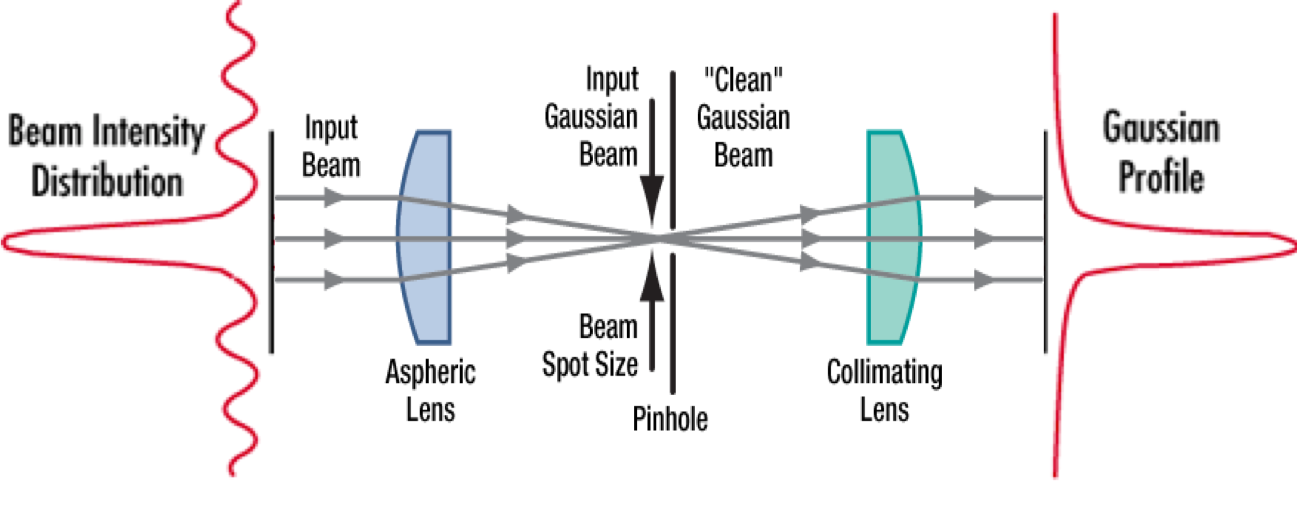}
\caption{A Gaussian beam with high-frequency aberrations is spatially filtered by a pinhole acting in the Fourier Transform plane of the lens.}
\label{fig:spatialfilter}
\end{center}
\end{figure}

Spurious diffractive effects can also be spatially filtered in the Fourier plane, or by applying a mask on the Fourier Transform in software to reconstruct only the image of interest. In the example shown in Fig.~\ref{fig:spatialfilter}, a pair of converging lenses separated by the sum of their focal lengths and an aperture stop can be used to improve the spatial quality of an imperfect Gaussian laser beam. A spatial filter uses the principle of Fourier optics to alter the structure of a beam of coherent light. Placing the pinhole aperture stop at the focus, the pinhole acts in the Fourier transform plane of the lens to eliminate structure with higher spatial frequencies, which produce light furthest from the central position. In practice a microscope objective and pinhole is typically used to remove aberrations and improve the quality of a Gaussian laser beam.

The first lecture concluded with a brief review of diffraction mitigation in beam instrumentation, exemplified by the coronagraph for beam halo monitoring that detects synchrotron radiation from the~LHC. The optical system utilises an opaque disk to block the beam core, however, the limited diameter of the objective lens creates unwanted diffraction, which overlays the halo. By adding the field lens to image the objective lens, the unwanted diffraction moves radially out. A Lyot stop is then used to block the diffraction, allowing only the LHC halo to be imaged. For further details, please see~\cite{bib:Goldbatt2016, bib:Mitsuhashi2017}.

\section{Lasers, Technologies and Setups}
\subsection{Lasers: fundamental principles}
\subsubsection{The first laser}
Conventional light sources emit incoherent light of multiple frequencies in all directions, which is not so useful for beam instrumentation. In 1958 Arthur Schawlow and Charles Townes laid down the theoretical framework for an \emph{optical maser}, which is now known as the \emph{laser} (\underline{l}ight \underline{a}mplification by the \underline{s}timulated \underline{e}mission of \underline{r}adiation).  Lasers emit almost monochromatic (depending on the linewidth), coherent, highly directional beams that are extremely useful for precise measurements. We shall review some examples of how lasers are applied in beam instrumentation in the next section, however, first let us examine the~fundamental principles of laser operation, which provides some insight into the important parameters to consider when selecting a laser for an application.

\begin{figure}[h]
\begin{center}
\includegraphics[width=0.49\textwidth]{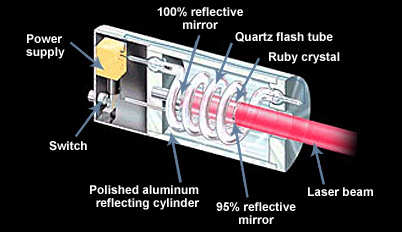}
\caption{Components of the first ruby laser~\cite{bib:firstlaser}.}
\label{fig:rubylaser}
\end{center}
\end{figure}

The first optical maser was built in 1960 by Theodore H. Maiman at Hughes Research Laboratories, and comprised the three essential components of a modern laser: an optical pump or excitation mechanism; the optical gain medium (a ruby crystal in this case); and an optical cavity resonator formed between mirrors surrounding the gain medium, as shown in Fig.~\ref{fig:rubylaser}.

\begin{figure}[h]
\begin{center}
\includegraphics[width=0.99\textwidth]{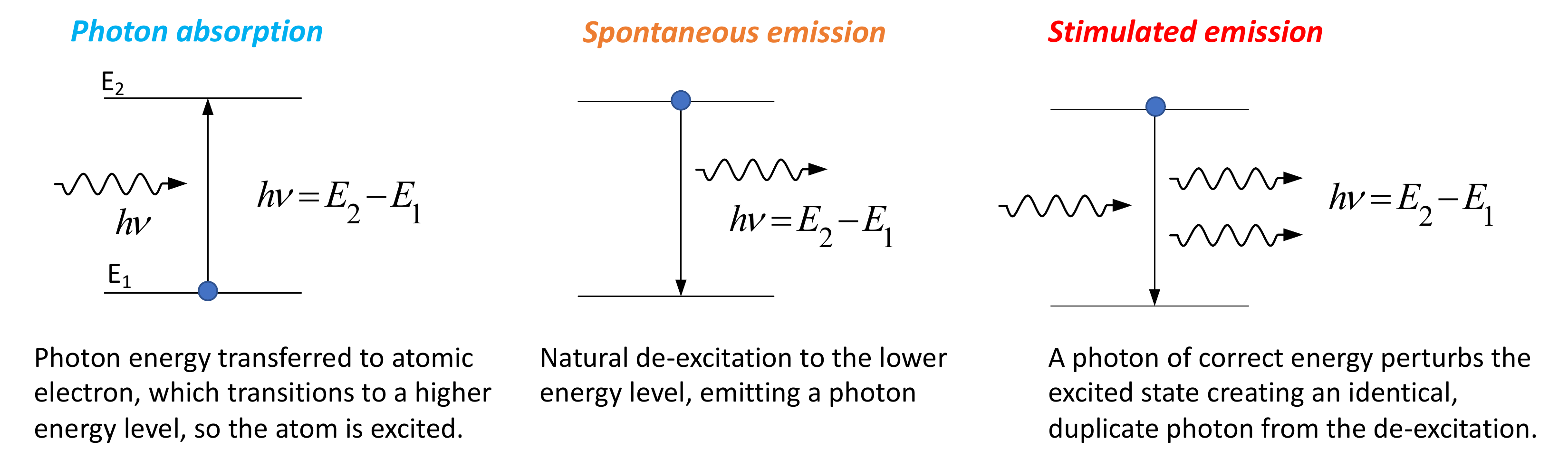}
\caption{Categories of transitions of electrons between atomic energy levels in a laser gain medium}
\label{fig:LaserTransitions}
\end{center}
\end{figure}

The physical operation of the laser is dependent on three types of atomic electron transition that are depicted in Fig.~\ref{fig:LaserTransitions}.

\begin{description}
\item[Photon absorption] occurs when an incident photon of the correct energy promotes an electron to a~higher energy level. The rate depends on the number of atoms in the lower energy level, the~incident photon flux, F and absorption cross section, $\sigma_{12}$.
\begin{equation}
\frac{dN_1}{dt} = - \sigma_{12} F N1 = - B_{12} N_1 u_\nu.
\end{equation}
\item[Spontaneous emission] occurs when the excited electron naturally falls back to the lower energy level, emitting a photon. The rate depends on the number of atoms in the higher energy level and lifetime, $\tau_{SP}$,
\begin{equation}
\frac{dN_2}{dt} = - \frac{N_2}{\tau_{SP}} = -A_{21} N_2.
\end{equation}
\item[Stimulated emission] occurs when an incident photon perturbs the excited electron, which then falls back to the lower energy level, emitting an identical duplicate photon. The rate depends on the~number of atoms in the higher energy level, the photon flux, $F$, and the stimulated emission cross section,
\begin{equation}
\frac{dN_2}{dt} = - \sigma_{21} F N_2 = -B_{21} N_2 u_\nu.
\end{equation}
\end{description}
The proportionality constants are the Einstein $A$ and $B$ coefficients, and $u_\nu$ is the energy density of radiation. We note that for a system in equilibrium, the absorption and emission processes must balance;
\begin{equation}
B_{12} N_1 u_\nu = A_{21} N_2 + B_{21} N_2 u_\nu.
\end{equation}
Solving for the energy density,
\begin{equation}\label{eqn:endensity}
u_\nu = \frac{N_2 A_{21}}{N_1 B_{12} - N_2 B_{21}} 
\end{equation}
The  Boltzmann distribution gives the probability that energy level $E_m$ in an arbitrary atom is occupied. When in thermal equilibrium, the relative population of levels is:
\begin{equation}
\frac{N_2}{N_1} = e^{-\frac{E_2 - E_1}{kT}}
\end{equation}
Substituting in Eq.~(\ref{eqn:endensity}) gives:
\begin{equation}
u_\nu = \frac{A_{21}}{B_{21}} \frac{1}{(B_{12}/B_{21}) e^{\frac{h\nu}{kT}}  - 1} 
\end{equation}
In order to agree with Planck's radiation formula, Einstein showed
\begin{equation}
B_{12} = B_{21}
\end{equation}
\begin{equation}
\frac{A_{21}}{B_{21}} = \frac{8 \pi h \nu^3}{c^3}
\end{equation}
Thus for atoms in thermal equilibrium, the ratio of stimulated to spontaneous emission rates is:
\begin{equation}
\frac{\mbox{stimulated emission}}{\mbox{spontaneous emission}} = \frac{B_{21} u_\nu}{A_{21}} = \frac{1}{e^{\frac{h\nu}{kT}} -1}
\end{equation}
Essentially, the rate of induced emission is extremely small at normal temperatures. Normal light sources are therefore dominated by spontaneous emission and produce incoherent light. 

\subsubsection{Population inversion}
\begin{figure}[h]
\begin{center}
\includegraphics[width=0.49\textwidth]{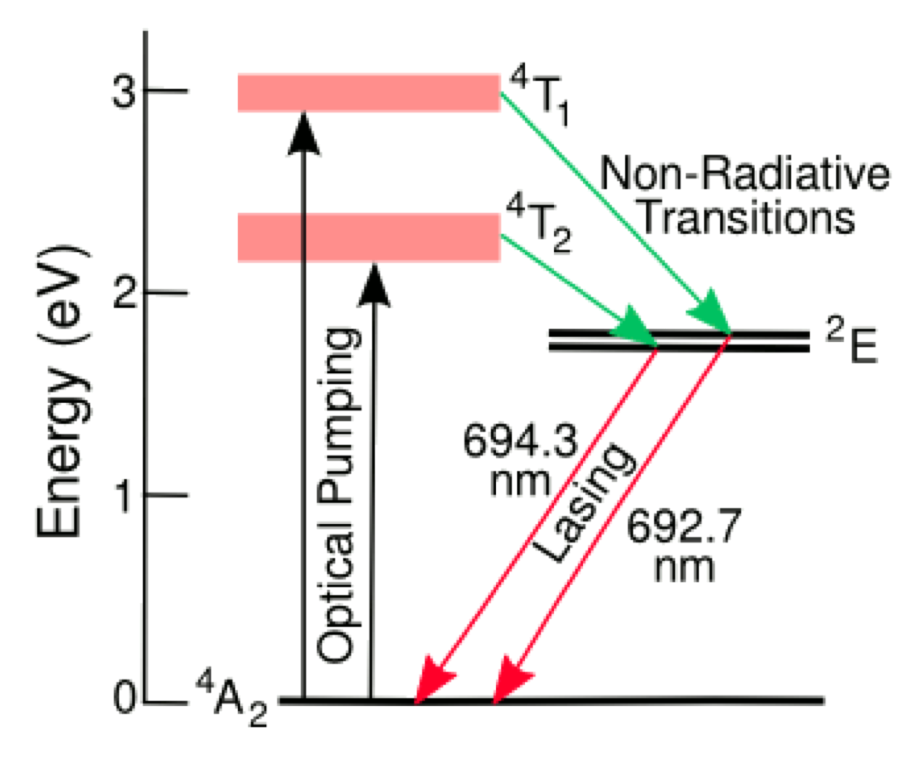}
\caption{Energy levels of the Cr$^{3+}$ doped ruby crystal in the first laser~\cite{bib:firstlaser}.}
\label{fig:rubylaserlevels}
\end{center}
\end{figure}

To create laser action by stimulated emission, more electrons must be placed in the upper energy level. This is known as \emph{population inversion} and is achieved by \emph{optical pumping}. In the Maiman's ruby laser the optical pump was a quartz flash tube that coiled around the ruby crystal in Fig.~\ref{fig:rubylaser}, and emitted an intense burst of light that excites the Cr$^{3+}$ dopants in the crystal. In this multi-energy level case, as in Fig.~\ref{fig:rubylaserlevels}, electrons are promoted from the ground state to the highest energy levels and then transition rapidly to populate the meta-stable energy level. Normally, more electrons populate the ground state than the meta-stable excited state, so absorption dominates over stimulated emission and there is no lasing.  When more Cr$^{3+}$ dopant ions have electrons promoted into the excited energy level, stimulated emission dominates over absorption and lasing occurs.

The initial photons are emitted by spontaneous emission in all directions, however, these photons stimulate emission from other Cr$^{3+}$, which gives rise to light amplification in the gain medium. Remarkably, each stimulated emission results in two photons that are identical to the incident photon, having the~same wavelength, phase, polarization and propagates in the same direction. The amplification is aided by multiple passes through the gain medium, due to recycling of the photons back and forth by mirrors at either end of the optical cavity. In fact this optical feedback mechanism creates an oscillator that resonates at the allow modes of the cavity, as described further below. One mirror is partially transmissive to allow some photons to escape the cavity as a narrow beam of highly directional, coherent laser light.

\subsection{Laser types and key parameters}
\subsubsection{Laser technologies}
Progress on laser development since the first laser was built in 1960 has been phenomenal, and there now exist a vast range of laser technologies to chose from. The scale and power of a laser can vary from the~pocket-sized laser-pointer of around 1\,mW based on a semiconductor diode laser, to the enormous 500 terawatt peak flash of light on target that is simultaneously generated from 192 beamlines in the~building-sized laser system that is the National Ignition Facility, Livermoore, US, which is used to study inertial confinement fusion.

\begin{sidewaysfigure}
\centering
\includegraphics[width=0.99\textheight]{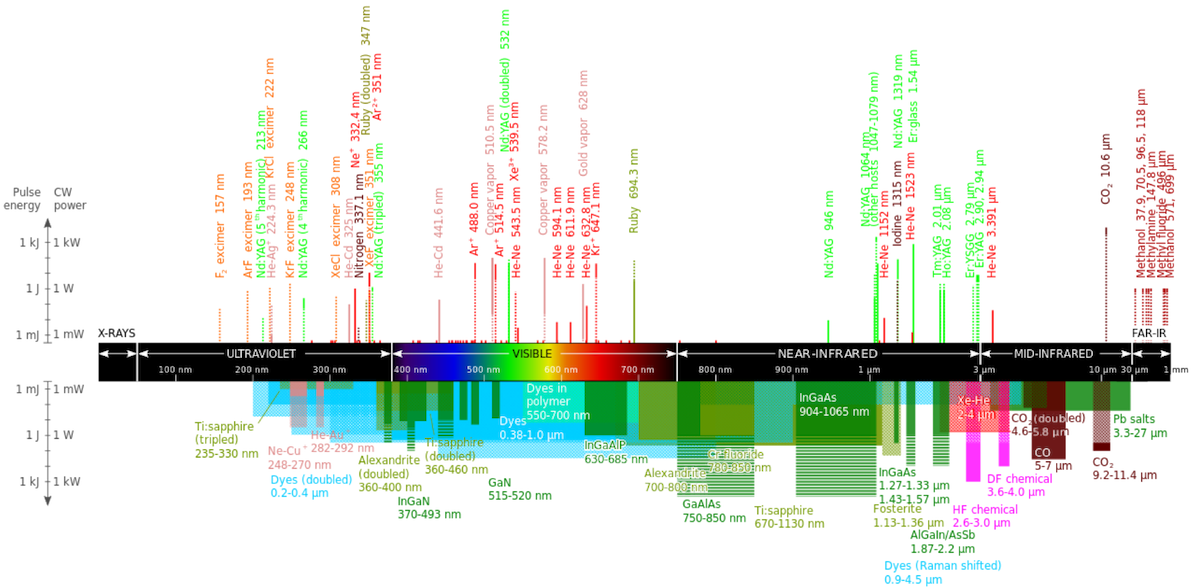}
\caption{Wavelengths of commercially available lasers~\cite{bib:LaserTypes}}
\label{fig:LaserTypes}
\end{sidewaysfigure}

When selecting a laser one of the first considerations is the choice of wavelength, which is defined by energy levels in the material of the optical gain medium. The technology choice also determines the~achievable optical power. A nice overview of commercially available laser wavelengths and powers is provided in Fig.~\ref{fig:LaserTypes}.

The main types of laser can be broadly categorised as listed below, which includes some examples of the gain medium:
\begin{itemize}
\item Gas lasers [HeNe, Argon, Krypton, CO$_2$]
\item Chemical lasers [COIL, AGIL, HF, DF]
\item Excimer lasers: chemical reaction involving excited dimer [F$_2$, ArF, KrF, XeCl, XeF]
\item Ion lasers: [Argon-Ion]
\item Metal-vapour lasers: [HeAg, NeCu, HeCd for UV wavelengths, etc], 
\item Solid state lasers [Ruby, Nd:YAG, Ti:sapphire]
\item Semiconductor lasers [GaN, InGaN, VCSELs]
\item Fibre lasers (Erbium doped)
\item Free electron laser
\end{itemize}
Further consideration must be given to the following laser parameters to appropriately match the requirements of the application:
\begin{itemize}
\item Pulse energy, or continuous wave (CW) power?
\item Fixed or tuneable wavelength? Required linewidth and spectral coherence?
\item Q-switched, repetition rate, mode-locked, master-oscillator power amplifier, free-space or fibre output?
\item Spatial beam quality, divergence, transverse modes, phase noise?
\end{itemize}
An important consideration is the optical power achievable as determined by the technology choice:
\begin{description}
\item[Continuous Wave] narrow linewidth, modest power; useful for interferometry. 1mW to <1kW. Continuous lasing.
\item[Q-switched] Pulse trains generated by electro-optic modulators within laser cavity. Pulse peak powers < a few 100 kW, in $\mu$s to ns pulses.
\item[Mode-locked] Short pulses generated by phase-locking cavity modes, as explained below, can generate pulse peak powers up to MW, for fs duration pulses.
\item[Chirp pulse amplification] A fast pulse is time-stretched before amplification and subsequently compression to attain pulse peak powers in the GW--PW regime, with pulse durations from ps to fs.
\end{description}

\subsubsection{Optical cavity modes, bandwidth and mode-locking}
\begin{figure}[h!]
\begin{center}
\includegraphics[width=0.59\textwidth]{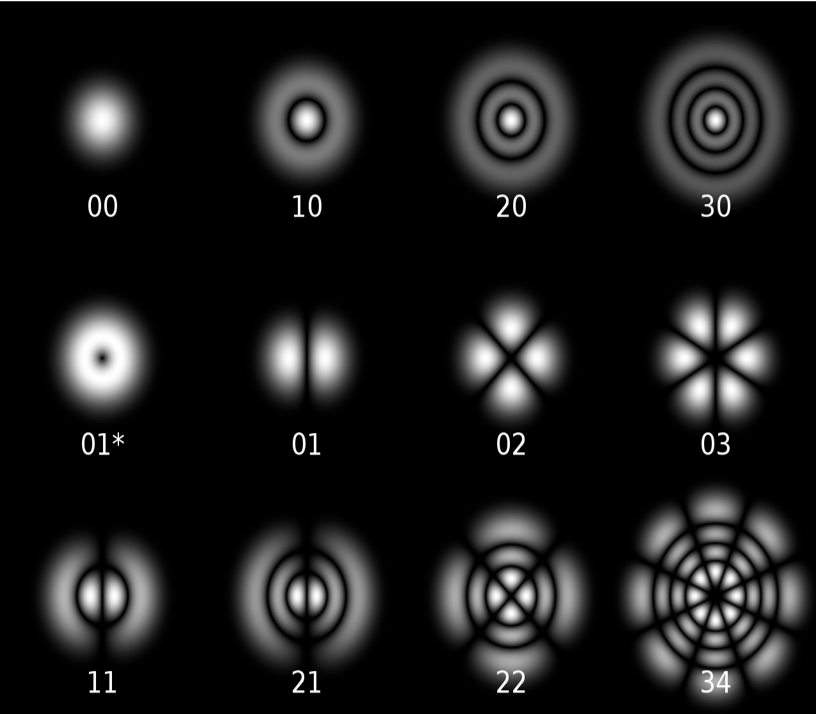}
\caption{Transverse Laguerre-Gauss TEM-NN modes of a cylindrical laser cavity.}
\label{fig:transversemodes}
\end{center}
\end{figure}

The two mirrors surrounding the laser gain medium create a Fabry-P\'e rot cavity, which determines the~properties of the master oscillator. This optical cavity enhances lasing only at certain resonant frequencies corresponding to longitudinal modes allowed by the cavity length, $L$, and mode number $n$.
The~electric field inside the cavity is
\begin{equation}
E(z,t)\Delta\omega = E_0 \cos(kz)\cos(\omega_L t).
\end{equation}
and applying the boundary conditions $E(L,t) = E(-L,t) = 0$ the allowed wavelengths are defined by $n\lambda = 2L$, the round trip length, which implies the resonant frequencies of the cavity modes are:
\begin{equation}
\omega = kc = \frac{2\pi c}{\lambda} = \frac{n \pi}{L}
\end{equation}
Thus the consecutive modes have a frequency difference,
\begin{equation}
\Delta\omega = \omega_{n+1} - \omega_n = \frac{\pi c}{L}.
\end{equation}
The transverse TEM-NN modes in a cylindrical cavity are the Laguerre-Gauss modes, as shown in Fig.~\ref{fig:transversemodes}. By placing a restrictive aperture in the cavity, the fundamental transverse TEM-00 mode is selected, resulting in a Gaussian output beam.

\begin{figure}[h!]
\begin{center}
\includegraphics[width=0.49\textwidth]{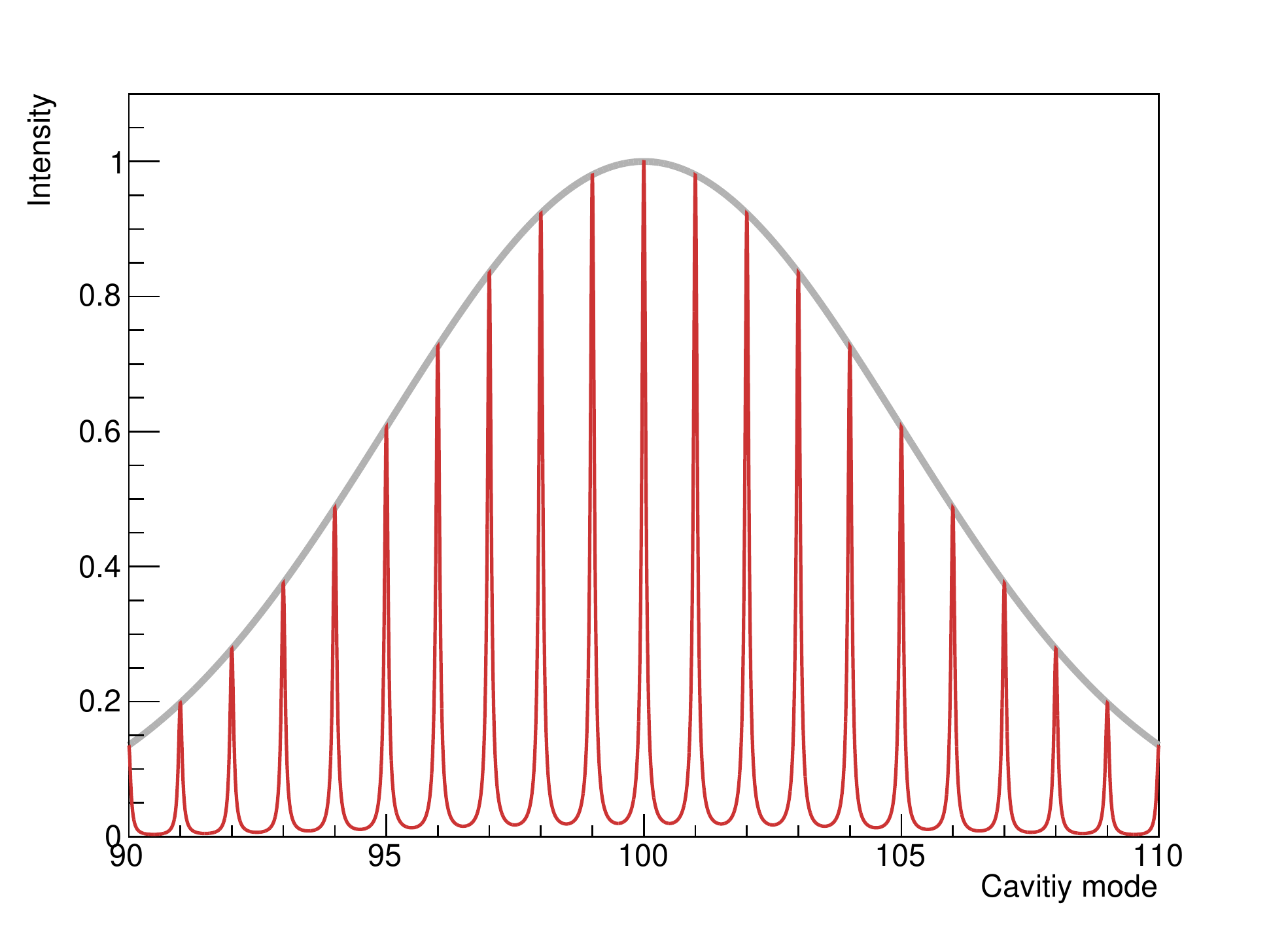}
\caption{Multiple longitudinal cavity modes are amplified within the gain envelope that is determined by the energy levels of the gain medium.}
\label{fig:cavitymodes}
\end{center}
\end{figure}

Although many lasers are nearly monochromatic, most do not emit at a single, pure frequency, but produce light with a natural bandwidth or range of frequencies. Primarily, the bandwidth is determined by energy levels of the gain medium and the corresponding range of frequencies that can be amplified, as in Fig.~\ref{fig:cavitymodes}. Within this range, the optical cavity length defines the frequency modes that are excited. Usually a laser will emit at multiple modes simultaneously, called \emph{multi-moded lasing}.

If the cavity modes are phase-locked then a temporal pulse can be generated in each round trip of the cavity, producing a repeating train of pulses. Ultrashort pulses implies keeping many modes in phase. The time-bandwidth product $TBP = \Delta \nu \Delta \tau$ for Gaussian and sech$^2$ pulses are transform-limited by their spectral content to:
\begin{align}
TBP_{\rm Gaussian} &= \frac{2 \ln 2}{\pi} \approx 0.441,\\
TBP_{\rm sech^2} &= \left ( \frac{2 \ln (1 + \sqrt{2}) }{\pi} \right  )^2 \approx 0.315.
\end{align}
This inverse relation between pulse duration and bandwidth, means that shorter duration pulses have the~largest frequency chirp. For example a sech$^2$ pulse of duration $1\,$ps implies a $\sim$1\,nm bandwidth, whereas a duration of $10\,$fs implies a $\sim$100\,nm bandwidth, which is a significant part of the visible spectrum. Further details on chirped pulses and their application for longitudinal bunch diagnostics can be found in A. Gillespie's contribution to these proceedings~\cite{bib:Gillespie2018}.

\subsection{Technologies and setups}
\subsubsection{Laser location, beam transport and synchronisation}
Lasers are typically sensitive and occasionally temperamental devices. When used for particle beam instrumentation are best kept in a safe laser cabin, typically in a surface building, away from the accelerator tunnel. This facilitates easy access to laser for maintenance or realignment, within the usual safety requirements for laser rooms: interlocks, safety shielding, goggles, and warning signs. A remote laser cabin reduces radiation exposure to both the laser and personnel, enabled a thermally stabilised environment to house the equipment, which may be mounted on pneumatically damped optical table to isolate from extraneous vibration.

The laser light must however be transported to (and sometimes from) the accelerator tunnel, for which there are two viable options. The first is a free space beam transport via series of mirrors with the light path running in protective tubes; this method has challenging beam pointing requirements over long distances, especially if tubes contain air, and are susceptible to thermally induced refractive index changes. This may be only option if very high optical power is required. A second option is the transport of light in optical fibres, which have the advantage of easy installation, though typically limit the peak power or pulse duration that can be successfully transported without pulse distortion due to non-linear effects in the fibre. Large mode area fibres have been successfully used to carry up to around a few kW peak power in laserwire systems at Linac4, and photonic crystal fibre has been used for higher peak powers with no discernible pulse distortion in the Petra-III laserwire system at DESY. In this system, if ~10\,ps laser pulses were to interact with the particle bunches the repetition rate of master oscillator needed to be carefully synchronised with the accelerator RF and with minimal timing jitter. Synchronisation was achieved by setting an external RF generator to a subharmonic of the accelerator RF frequency and comparing the phase between the laser pulse train and external RF. A feedback loop controlled a finely adjustable mirror within the laser cavity, which modified the cavity length to change the repetition rate, until a phase lock with the RF source was achieved. Finally lock the phase between the main clock to a~low noise (10MHz) reference from the accelerator RF timing.

\subsubsection{Light distribution and fibre-based multiplexing}
\begin{figure}[h!]
\begin{center}
\includegraphics[width=0.49\textwidth]{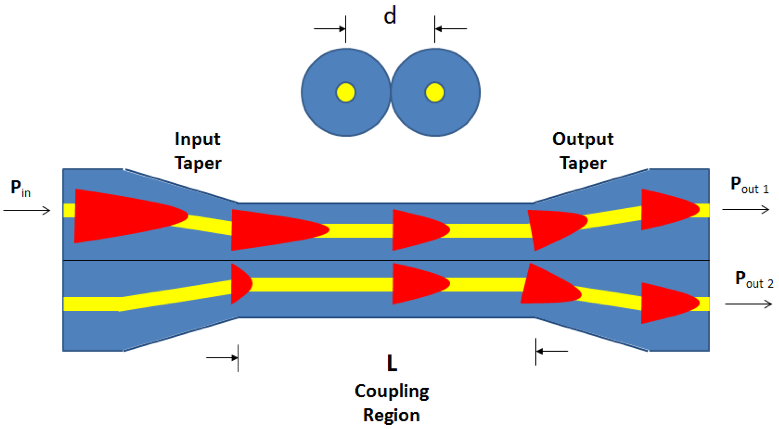}
\includegraphics[width=0.49\textwidth]{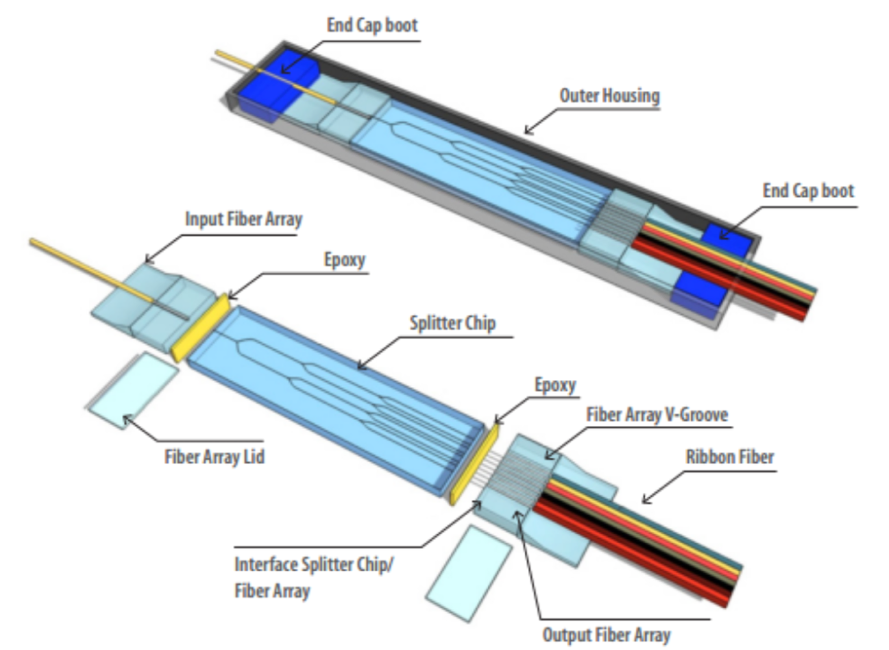}
\caption{Fibre-splitter components: a 2x2 fused biconic tapered coupler and 1x8 planar lightwave circuit}
\label{fig:FBTPLC}
\vspace{-30pt}
\end{center}
\end{figure}

Efficiencies in overall cost can often be achieved by the distribution of light from a single laser source to multiple beam instruments. This is typically achieved with beam-splitters in the case of free-space transport, or for optical fibre transport, the equivalent traditional method is to use a series of fused biconic tapered (FBT) couplers, that each split fibre-coupled light from one input to two output fibre channels. Such devices are manufactured by essentially fusing two twisted single fibres together as they are elongated, so that the light incident on one fibre is shared between the two outputs, in a split ratio that can be precisely controlled by adjustment to the fused geometry. Alternatively a tree of one to many (2N) waveguide splitters can be created on planar lightwave circuit chips, as shown in Fig.~\ref{fig:FBTPLC}.

\begin{figure}[h!]
\begin{center}
\includegraphics[width=0.64\textwidth]{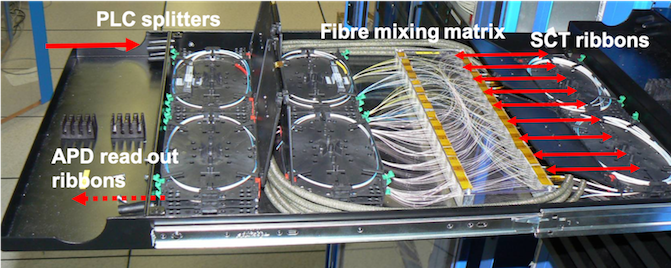}
\includegraphics[width=0.35\textwidth]{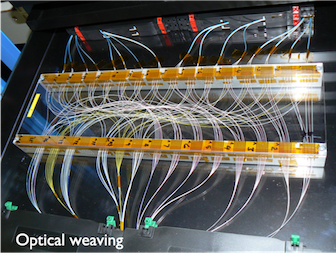}
\caption{This rack-mounted 1U tray represents one ninth of the FSI fibre-splitter tree, built to distribute fibre-coupled laser light to and from 842 interferometers inside ATLAS.}
\label{fig:FSItree}
\vspace{-20pt}
\end{center}
\end{figure}

As an example, the ATLAS Frequency Scanning Interferometry system was built to distribute light from a remote laser housed in a surface building to 842 on-detector interferometers via a optical fibre-splitter tree made from a combination of PLC and FBT splitters, as shown in Fig.~\ref{fig:FSItree}. The 842 return signal fibres were also hand-weaved through the same network of fibres and fusion spliced to ribbons that went to multiplexed readout boards. Nowadays, such optical weaving can be created commercially on flexible substrates for high-fibre-count cross-connect systems.

Other useful technologies are fast optical fibre switches that enable active control of the light distribution network rather than fixed, passive sharing of light. Such switches are typically based on MEMS technology and can carry moderate CW power, albeit with switching speeds limited to around 1\,ns. Ultra-high bandwidth signals can be transmitted optically by electro-optic modulators, which convert electrical signals using waveguide based Mach-Zehnder interferometers based on electro-optic effect. This enables high bandwidth transmission from the accelerator to a remote detector over fibre.

\subsection{Summary}
In this section we reviewed the fundamental operational principles of a laser system, the various types of laser commercially available and have identified some of the important parameters to consider when selecting a laser.  We have seen that when developing laser-based instrumentation for an accelerator environment consideration must be given to the safe access for personnel and the radiation tolerance of laser equipment, which normally implies a remotely housed laser with optical beam transport from the~laser to the accelerator. Various fibre-based technologies for light distribution were outlined. In the~next section we turn to applications of lasers in beam instrumentation and consider how the intrinsic properties of laser light have been exploited to date in the design of optical beam instrumentation.

\section{Laser-based Beam Instrumentation}
\subsection{Introduction}
In the above sections we have seen that lasers produce monochromatic, coherent and highly directional light that can be focused to sub-micron scales; properties that are extremely useful for state-of-the-art beam diagnostics. This section examines a selection of beam instrumentation applications, with an emphasis on non-invasive methods in which the particle beam interacts directly with laser generated photons as identified by item (b) in Section~\ref{sec:opticsintro}. Examples of item (e), electro-optic conversion of the~beam signal, can be found in~\cite{bib:Gillespie2018}.

\subsection{Laserwires}
\subsubsection{Motivation}
A reliable instrument to measure emittance at circular accelerators is the wire scanner. When a thin wire is rapidly scanned through a particle beam its transverse profile can be reconstructed as a function of the~wire position, either by measuring the secondary emission current on the wire, or by recording the~flux of secondary particles created as the beam interacts with the wire. In recent years, wire-scanners have been developed to be nearly non-destructive and can reach spatial resolutions down to a few microns~\cite{bib:Sirvent2017}. Nonetheless, the wire diameter cannot be reduced below the limit required for tensile strength. When the bunch charge density is sufficiently large, the deposited energy can lead to sublimation damage even of carbon wires. In high intensity beam scenarios where the wire would not survive, non-invasive techniques are an attractive alternative. Replacing the mechanical wire with a narrow laser beam eliminates the~possibility for the particle beam to directly damage the probe instrument. Such \emph{laserwires} can also exploit the ability to focus light to the micron-scale, making them particularly suited to lepton accelerators where the transverse particle beam is extremely small.

\subsubsection{Electron laserwires} \label{sec:electronlaserwires}
\begin{figure}[h!]
\begin{center}
\includegraphics[width=0.6\textwidth]{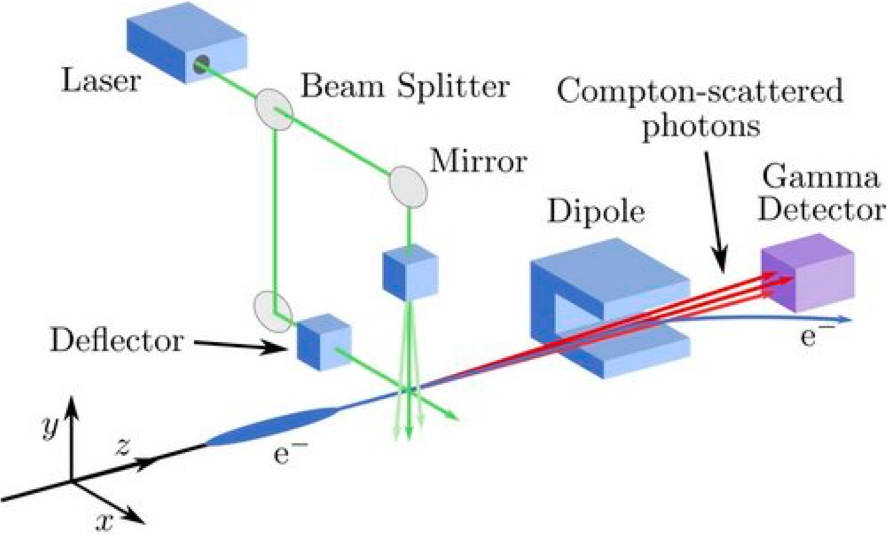}
\caption{\small Principle of laserwires for electron beams~\cite{bib:Nevay2014a}.}
\label{fig:electronlaserwire}
\end{center}
\vspace{-10pt}
\end{figure}

A \emph{laserwire} operates by shining a narrow beam of laser light across a particle beam and measuring the~interaction rate as the laser beam position is scanned in the transverse plane. For an electron beam, the photon interacts by (inverse) Compton scattering and the forward scattered photons are recorded by a downstream detector, while the electron beam is deflected by a dipole magnet, as shown in~Fig.~\ref{fig:electronlaserwire}.

As the narrow laserwire is scanned across the particle beam, the charge density in the electron beam can be reconstructed by plotting the intensity at the detector versus the laserwire position.
The~beam profile in the orthogonal transverse dimension can be obtained simply by arranging for a~second, orthogonally incident laserwire. Alternatively, with careful design of the focusing optics, a longitudinal scan of a tightly focused laser spot can be deconvolved to recover the electron beam profile.

\subsubsubsection{Inverse Compton Scattering}
When a laserwire at optical wavelengths interacts with an ultra-relativistic electron  beam, the electrons lose energy while the photons gain energy, in contrast to the standard Compton effect, hence this process is called \emph{inverse} Compton scattering. The energy $\hbar \omega_0$ of an incident photon in the laboratory frame $S$ is boosted in the electron frame $S^{\prime}$ by the relativistic Doppler shift,
\begin{equation}
\hbar \omega^{\prime} = \gamma \hbar \omega_0 (1 - \beta \cos \theta), \label{eqn:DopplerShift}
\end{equation}
where $\theta$ is the angle in $S$ between the propagation directions of the incident photon and electron, with $\beta = v/c$. Furthermore, the time interval between the arrival of photons from the direction $\theta$ is shorter by a factor $\gamma (1 - \beta \cos \theta)$ in $S^{\prime}$ compared to $S$, with a corresponding increase in the arrival rate of photons and number density. The energy density of radiation in $S^{\prime}$ is therefore
\begin{equation}
U^{\prime} =  [\gamma (1 - \beta \cos \theta)]^2 U
\end{equation}
In the simple case that the laserwire is orthogonally incident to the electron beam in $S$ then $U^{\prime} =  \gamma^2 U$. Provided that $\hbar \omega \ll m_e c^2$, the interaction in $S^{\prime}$ frame is just Thomson scattering, with cross-section $\sigma_T = 6.65\times 10^{-25}$\,cm$^2$, hence the rate of energy loss of the electron is the rate at which energy is reradiated. The energy loss rate is invariant in inertial frames, so we have
\begin{equation}
\frac{dE}{dt} = \frac{dE}{dt}^{\prime} = \sigma_T c U^{\prime} =\sigma_T c U \gamma^2
\end{equation}
The rate of energy gain to the photons in $S$ is therefore
\begin{align}
\frac{dE}{dt} & = \sigma_T c U \gamma^2 - \sigma_T c U \\
		    &=  \sigma_T c U (\gamma^2 - 1)\\
		    &= \sigma_T c U \beta^2 \gamma^2
\end{align}
We identify that the average energy of the scattered photons in the laboratory frame is:
\begin{equation}
\hbar \omega =  \beta^2 \gamma^2 \hbar \omega_0  \approx \gamma^2 \hbar \omega_0
\end{equation}
The factor of $\gamma^2 = \frac{1}{1-\beta^2}$ indicates the photons are scattered by the ultra-relativistic electron to large energies, with an energy spectrum given by~\cite{bib:Shintake1999, bib:Arutyunian1963}, 
\begin{equation}
\frac{d\sigma_{IC}}{d w} = \frac{3  \sigma_T}{8 \epsilon_1} \left [ \frac{1}{1-w} + 1 - w + \left [ \frac{w}{\epsilon_1 (1-w)} \right ]^2 - \frac{2 w}{\epsilon_1 (1 - w)} \right] ,
\end{equation}
where $\epsilon_1 = \gamma \hbar \omega_0 / m_e c^2$ is the normalized energy of the incident photon in the electron rest frame $S^{\prime}$, and $w = \hbar \omega / E_e$ is the normalised energy of the emitted photon. The scattered radiation is confined within a cone of a half angle that is a few times the critical angle,
\begin{equation}
\alpha_C = \frac{\sqrt{1+2\epsilon_1}}{\gamma}.
\end{equation}
The electrons that interact are significantly degraded in energy and in the subsequent accelerator lattice can be recorded as beam losses.

The differential cross-section of Compton scattering is described by the Klein-Nishina formula~\cite{bib:KleinNishina1929} and scales inversely with the squared mass of the particle, hence is the interaction rates are significant only for light particle beams. Even for typical parameters of lepton beams, high power, pulsed lasers are required to generate sufficient photon flux to be measurable for diagnostics purposes. 

\begin{figure}[h!]
\begin{center}
\includegraphics[width=0.6\textwidth]{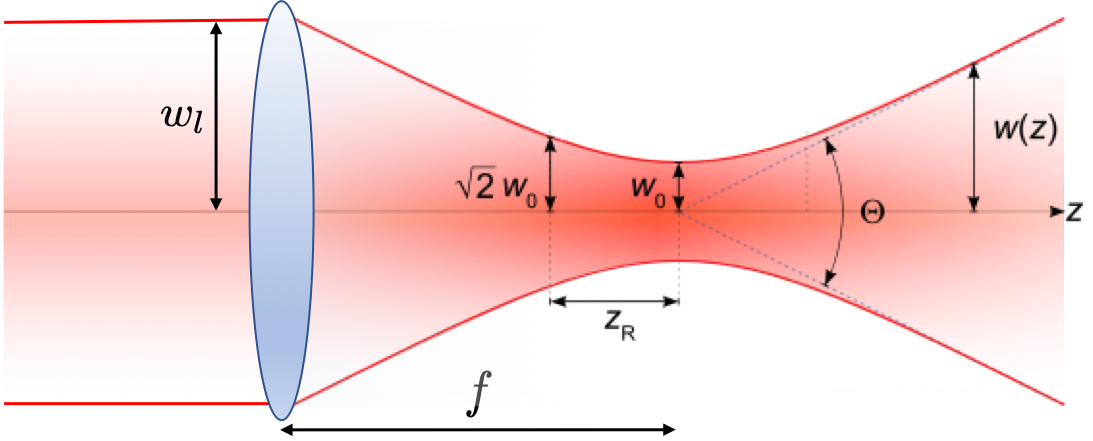}
\caption{\small The focus of a Gaussian laser beam with definitions for the $1/e^2$ irradiance contour (not to scale).}
\label{fig:GaussianBeam}
\end{center}
\end{figure}

\subsubsubsection{Gaussian beams}
The laserwire geometry is defined by Gaussian beam optics, as shown in Fig.~\ref{fig:GaussianBeam}. The radius at position $z$ along the beam is the hyperbolic contour,
\begin{equation}
w(z) = w_o\sqrt{1+\left (\frac{z - z_0}{Z_R}\right )^2},
\end{equation}
defined where the intensity drops to $1/e^2$ of the peak value at that position ($w(z) = {\rm FWHM}(z) / \sqrt{2 ln2}$\,). The laser beam is focused to beam waist $w_0$ that must be smaller than or comparable in size to the electron beam;
\begin{equation}
w_0 =\frac{M^2 f \lambda}{ w_l \pi},\label{eqn:laserwaist}
\end{equation}
where $M^2$ is a measure of the quality of the transverse mode (a value of $M^2=1$ would be an ideal Gaussian), and $w_l$ is the beam radius at the lens of focal length $f$.
The Rayleigh range $z_R$ is the propagation distance over which the beam size grows to $\sqrt2$ of the size at the laser waist $w_0$,
\begin{equation}
Z_R = \frac{\pi w_0^2}{M^2 \lambda}.
\end{equation}

\begin{figure}[h!]
\begin{center}
\includegraphics[width=0.9\textwidth]{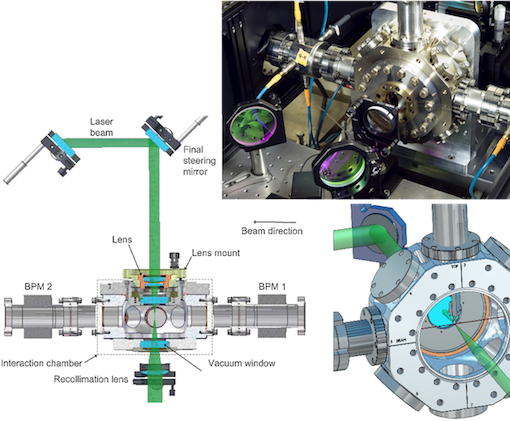} 
\caption{\small Laserwire interaction chamber at the ATF beamline and CAD model showing laser focus (side flange removed)~\cite{bib:Boogert2010}.}
\label{fig:ATFlaserwire}
\end{center}
\end{figure}

\subsubsubsection{Electron laserwire design}
The focal spot size is a key parameter for the resolution of an electron beam laserwire and achieving a~movable, micron-scale focus demands a rigorous opto-mechanical design. From Eq.~(\ref{eqn:laserwaist}), the smallest focal spot size is achieved when an expanded laser beam strikes the final focusing optics as close as possible to the interaction point. Light focused into the interaction chamber through the vacuum window must pass through carefully designed transmissive optics that deliver a beam with minimal aberrations. A series of lenses is necessary, including an aspheric surface to correct for the spherical aberrations that would otherwise degrade the beam quality, increase $M^2$ and thus enlarge the achievable spot size. The~vacuum window forms an integral part of the lens and must withstand the pressure difference with minimal deformation. The optics are anti-reflection coated to eliminate ghost images, must avoid excessive energy absorption from the high power laser, and be radiation tolerant, hence the lens material is typically fused-silica. An example interaction chamber mounted on the ATF beamline at KEK is shown in Fig.~\ref{fig:ATFlaserwire}, which was capable of measuring electron beam sizes down to 4.8\,$\mu$m~\cite{bib:Boogert2010}.

\begin{figure}[h!]
\begin{center}
\includegraphics[width=0.49\textwidth]{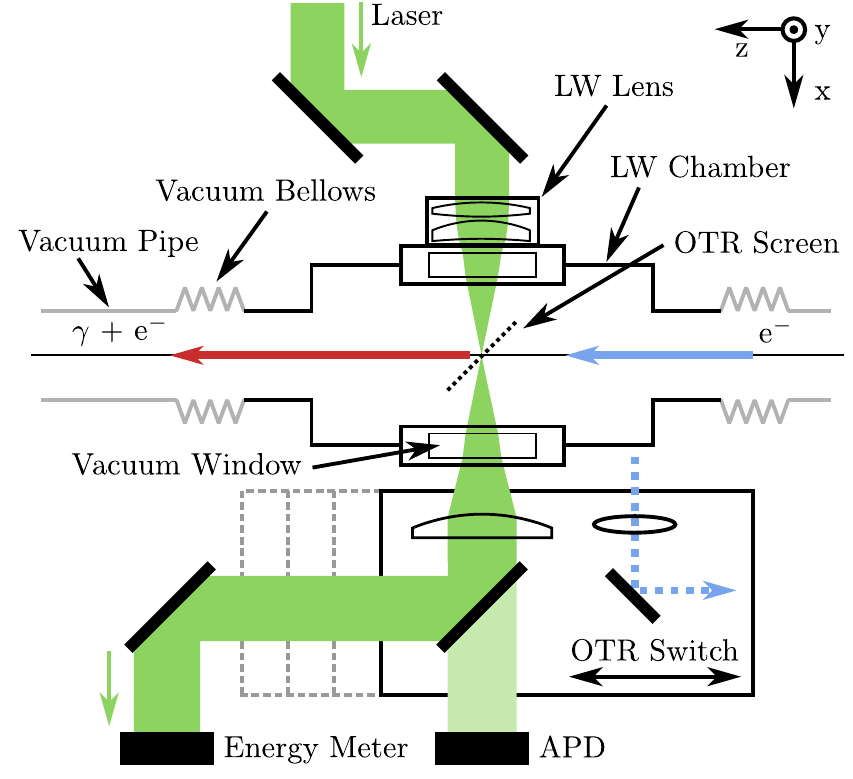} 
\includegraphics[width=0.49\textwidth]{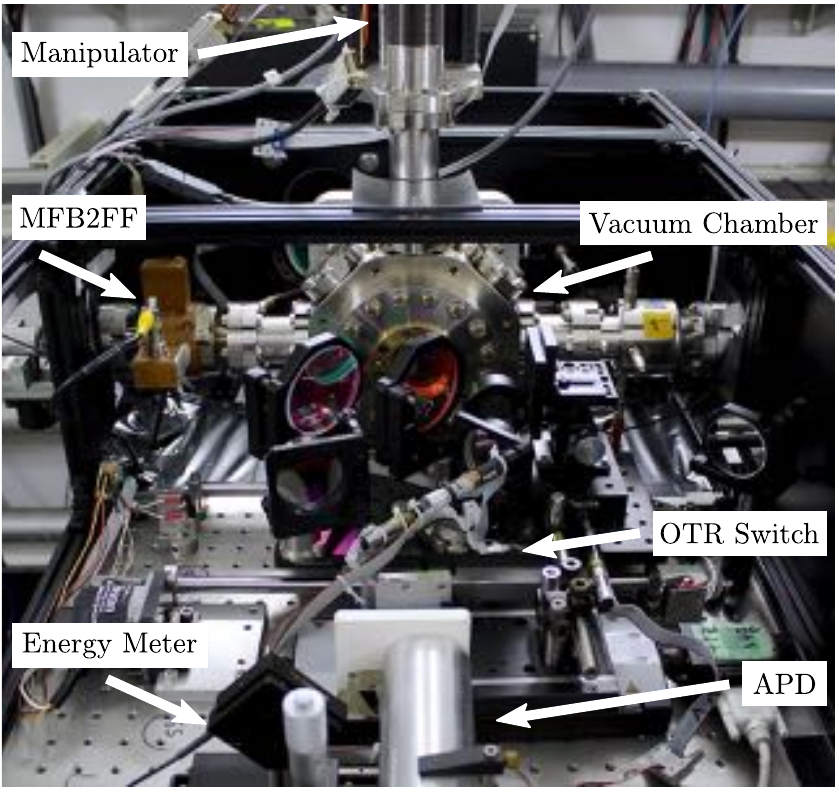} 
\includegraphics[width=0.50\textwidth]{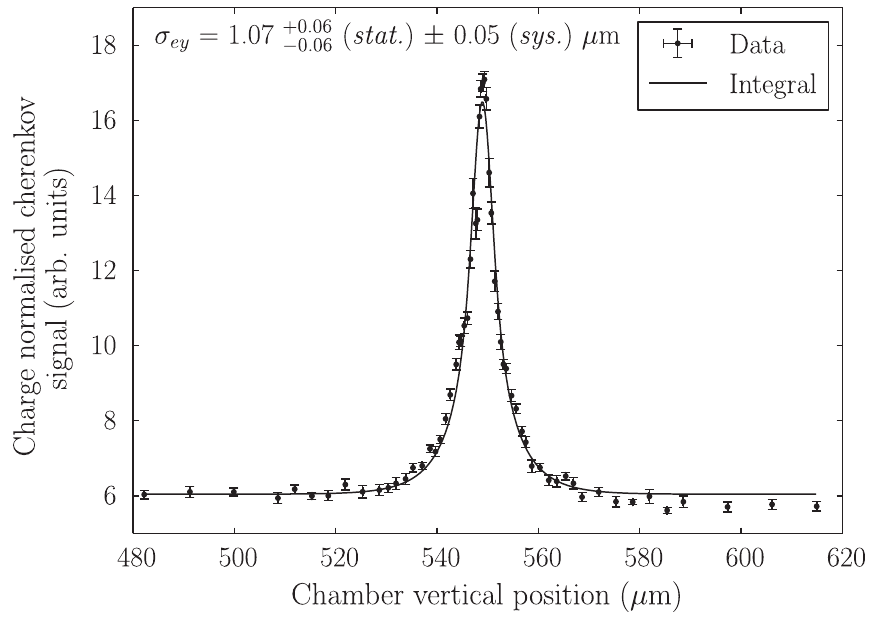} 
\includegraphics[width=0.48\textwidth]{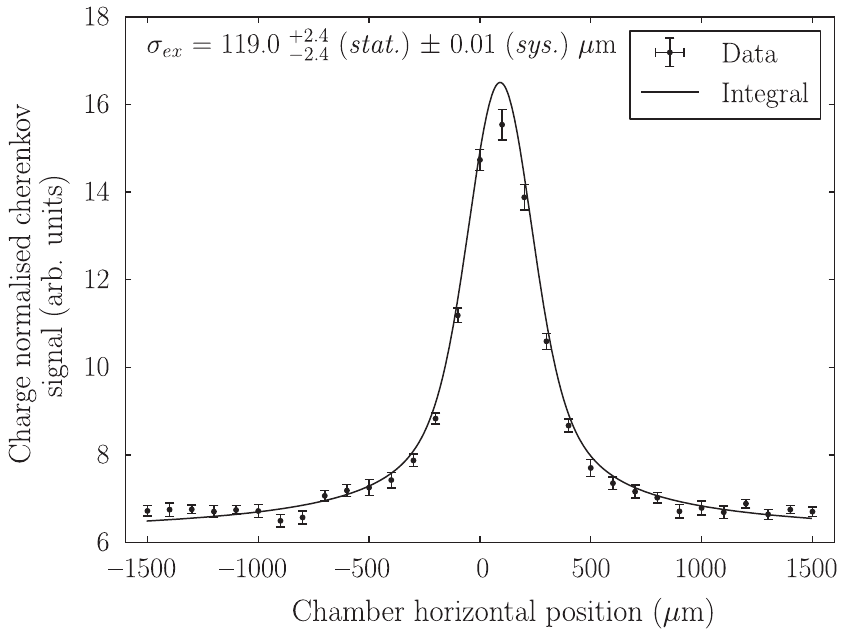} 
\caption{\small Laserwire configuration and installation at ATF2 (including a high resolution MFB2FF cavityBPM) and the measured electron beam profiles in the vertical and horizontal directions~\cite{bib:Nevay2014b}.}
\label{fig:ATF2laserwire}
\end{center}
\end{figure}

The laserwire was upgraded at the ATF2 facility, as shown in Fig.~\ref{fig:ATF2laserwire},
with the aim of measuring the electron beam profile with an aspect ratio of $1\times 120 \,\mu$m. The highly asymmetric beam presents a~challenge because in order to achieve a small enough focal spot to resolve the $1\,\mu$m electron beam in the~vertical direction, the Rayleigh range of the laser focus is 15~$\mu$m for the visible wavelength $\lambda= 532$\,nm laser. Therefore the vertical size of the laser focal spot varies significantly across the~horizontal dimension of the beam. This creates a non-Gaussian detector response during a vertical scan of the~laserwire, due to the shape overlap with the electron beam in the wings of the distribution. Precise laser characterization and horizontal laserwire scans allowed a detailed model of the overlap integral~\cite{bib:Agapov2007} to be applied, resulting in a successful measurements of both the horizontal and vertical electron beam sizes~\cite{bib:Nevay2014b}.

Several international laboratories have developed electron laserwires, including a UV laser-based beam profile monitor at the Stanford Linear Collider~\cite{bib:Alley1996}, a two-dimensional laserwire at PETRA-III in DESY~\cite{bib:Bosco2008}, and the laserwire at the Beijing Electron-Positron Collider II~\cite{bib:He2015}.

\subsubsection{H$^{-}$ laserwires}
Hydrogen ion accelerators form the front end of high power proton drivers for numerous applications, including next generation spallation neutron sources, neutrino beams, a future muon collider and accelerator-driven reactors for the transmutation of nuclear waste. The beam powers generated are in the~megawatt regime and beam currents typically exceed 10\,mA, which is above the damage threshold for conventional interceptive beam diagnostics. A laserwire offers a non-invasive probe to measure beam profiles and emittance at hydrogen ion accelerators. In contrast to the small electron beams described above, the transverse size of hydrogen beams is much larger, typically at the mm level, only requiring the laserwire to be focused to  $<100\,\mu$m and the corresponding Rayleigh range is typically longer than the transverse dimension of the bunch, which simplifies the analysis.

Hydrogen ion laserwires were originally built at Los Alamos National Laboratory~\cite{bib:Cottingame1985, bib:Connolly1992, bib:Yuan1993}, and developed further at facilities including the Brookhaven National Laboratory LINAC~\cite{bib:Connolly2011}, the Spallation Neutron Source at Oak Ridge National Laboratory~\cite{bib:Liu2010, bib:Huang2013, bib:Liu2013}, CERN's LINAC4~\cite{bib:Gibson2013, bib:Hofmann2015, bib:Hofmann2016, bib:HofmannThesis, bib:Hofmann2018a, bib:Hofmann2018b} and J-PARC~\cite{bib:Miura2016}. 

\begin{figure}[h!]
\begin{center}
\includegraphics[width=0.6\textwidth]{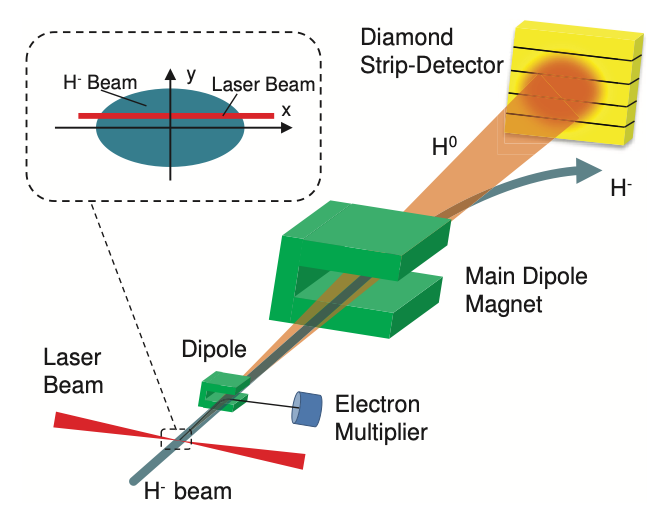} 
\caption{\small Principle of laserwire for $H^{-}$ beams~\cite{bib:Hofmann2018b}.}
\label{fig:Hminuslaserwire}
\end{center}
\end{figure}

A hydrogen ion laserwire operates on the principle of photo-detachment, in which an incident photon has sufficient energy ($>0.756$\,eV) in the rest frame of the $H^{-}$ ion to permanently eject the~weakly bound outer electron from the negative ion. The result is a low energy electron and a neutralised hydrogen atom that essentially continues in the original direction of the ion.
\begin{equation}
H^{-} + \gamma \rightarrow H^0 + e^{-}
\end{equation}
Either the electron or the neutralised hydrogen, or preferably both, are recorded by appropriate downstream detectors, as shown in Fig.~\ref{fig:Hminuslaserwire}.
In diagnostics applications, only a small fraction of the beam is neutralised, and the remaining $H^{-}$ beam is deflected by a dipole magnet. The low energy electrons can also be deflected by small dipole magnet to be captured by a charge sensitive detector such as a Faraday cup, a small diamond detector, or an electron multiplier. 

\begin{figure}[h!]
\begin{center}
\includegraphics[width=0.6\textwidth]{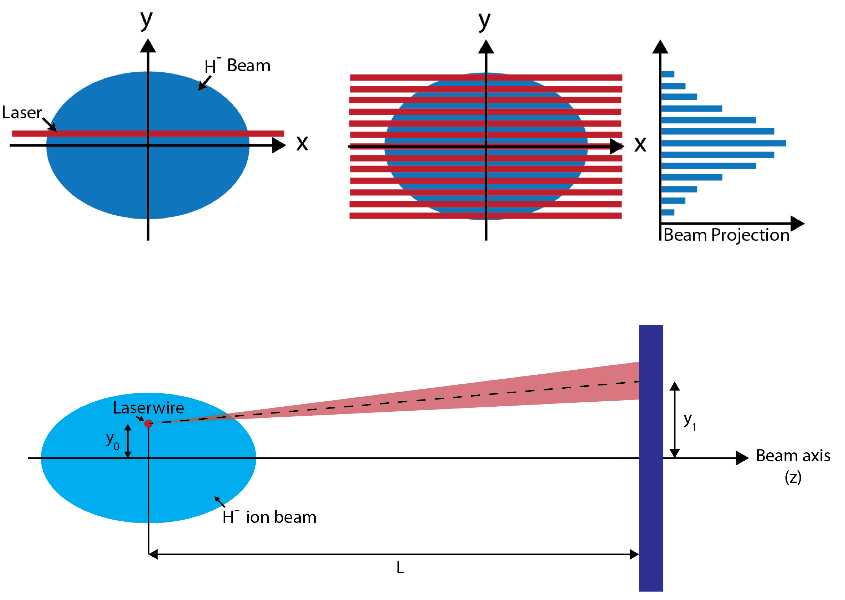} 
\caption{\small Schematic of transverse beam profile reconstruction and the sensitivity to angle $\frac{y_1 - y_0}{L}$ at the detector distant $L$ from the $H^{-}$ laserwire.}
\label{fig:emittanceprinciple}
\end{center}
\end{figure}

As for an electron laserwire, by scanning the laserwire position the beam profile can be reconstructed from the detected intensity of either decay product.  Moreover, the transverse emittance can be reconstructed by analysing the distribution of the beamlet of neutralised H$^{0}$ particles after they drift to the spatially sensitive downstream detector. They arrive unperturbed by the magnet field and so retain the angular information from the interaction point, as shown in Fig.~\ref{fig:emittanceprinciple}, hence the transverse phase space of the beam can be plotted to determine the beam emittance.

\begin{figure}[h!]
\begin{center}
\includegraphics[width=0.5\textwidth]{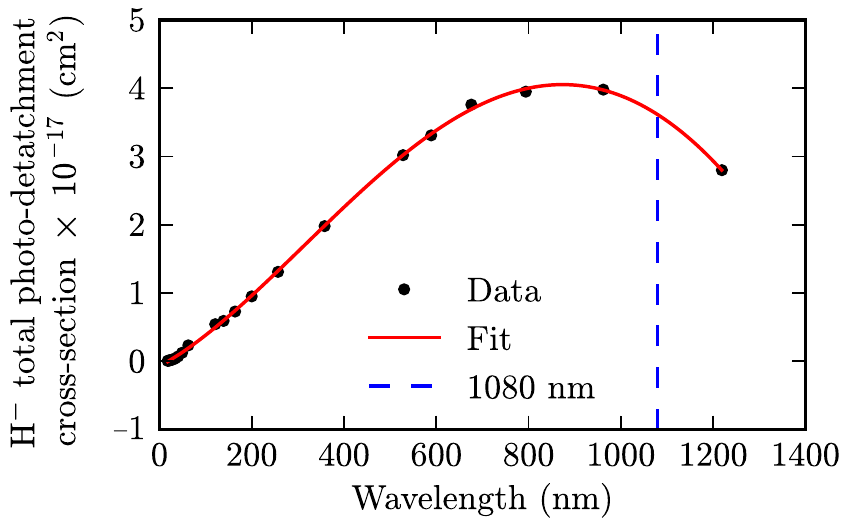} 
\caption{\small Photo-detachment cross-section of a non-relativistic $H^{-}$ ion~\cite{bib:Hofmann2015} with data from~\cite{bib:BroadReinhart1976} (Feshbach resonance not shown).}
\label{fig:photodetachcrosssection}
\end{center}
\vspace{-20pt}
\end{figure}

The photo-detachment cross-section depends on the Doppler shifted energy of the photon in the~rest frame of the ion, which is identical in form to Eq.~(\ref{eqn:DopplerShift}). The cross-section peaks at a photon wavelength close to 900\,nm \cite{bib:BroadReinhart1976}, so depending on the H$^{-}$ beam energy, the laser wavelength is selected such that the Doppler shift brings the photon energy into the region of the peak, as in Fig.~\ref{fig:photodetachcrosssection}. For example at CERN's LINAC4, a 1080\,nm wavelength laser was selected for the 160\,MeV H$^{-}$ beam, which is Doppler shifted close to the peak~\cite{bib:Hofmann2015}.

\begin{figure}[b!]
\begin{center}
\includegraphics[width=0.99\textwidth]{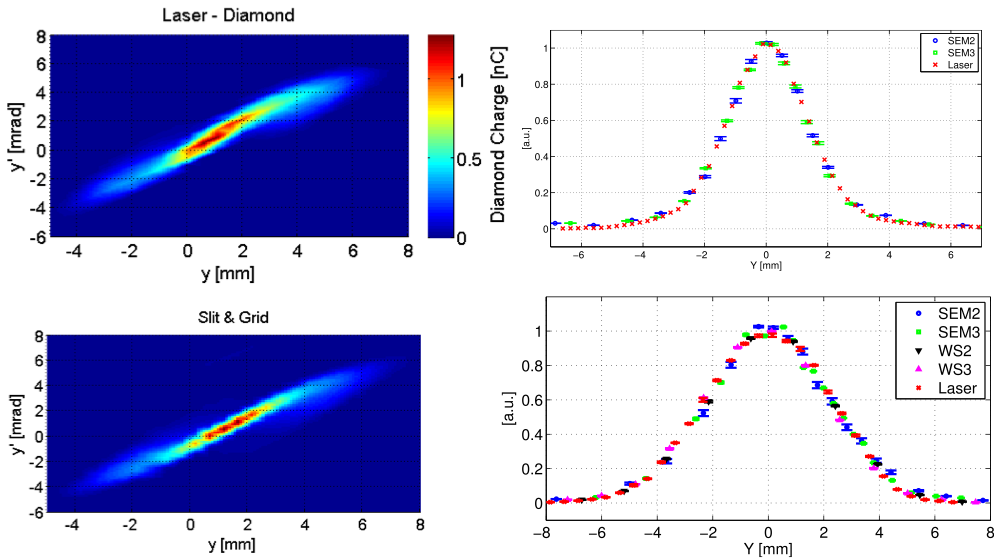} 
\caption{\small Comparison between $H^{-}$ laserwire and conventional diagnostics at CERN's LINAC4. Emittance measurements shown at 12 MeV, and beam profiles at 80 MeV (upper) and 107 MeV (lower), for SEM-grid and wirescanner (WS) and laserwire measurements.~\cite{bib:Hofmann2016, bib:Hofmann2018a}}
\label{fig:L4laserwireresults}
\end{center}
\vspace{-10pt}
\end{figure}

\begin{figure}[t!]
\begin{center}
\includegraphics[width=0.9\textwidth]{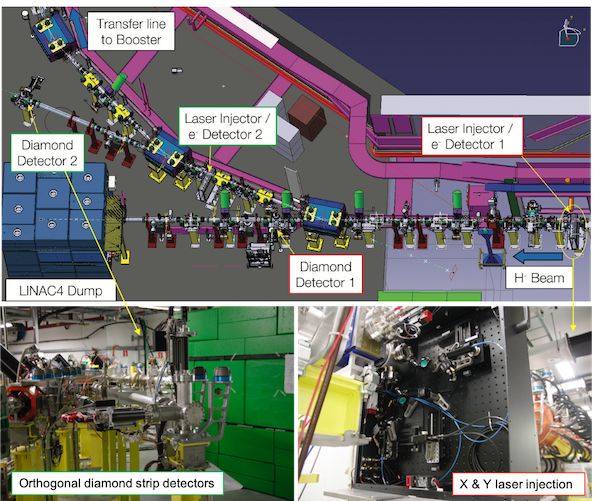} 
\caption{\small Dual-axis, dual-station $H^{-}$ laserwire installed at CERN's LINAC4.~\cite{bib:Hofmann2018b}}
\label{fig:L4laserwirelayout}
\end{center}
\vspace{-10pt}
\end{figure}

Importantly, the photo-detachment cross-section is seven orders of magnitude higher than for inverse Compton scattering, which considerably eases the peak power and pulse energy requirements for the laser. A low-power (<kW peak) fibre coupled laserwire prototype was demonstrated during the~commissioning stages of CERN LINAC4 at 3\,MeV~\cite{bib:Hofmann2015}, 12\,MeV~\cite{bib:Hofmann2016} 50/80/107\,MeV~\cite{bib:Hofmann2018a}. A sharp $H^{0}$ signal corresponding to each laser pulse  was observed at the detector. The signal was well above the~residual gas-stripping background from the upstream linac, which arrived spread out in time, and was therefore easily distinguished and suppressed with respect to the signal pulses, by using a diamond detector with a fast ($\sim$ns) response. The results shown in Fig.~\ref{fig:L4laserwireresults}
were found to be agree to within 2\% of conventional beam diagnostics. A permanent LINAC4 laserwire system was recently commissioned at 160\,MeV~\cite{bib:HofmannThesis, bib:Hofmann2018b}, comprising  two stations of dual-axis laserwires, readout by multi-channel orthogonal diamond strip detectors, placed downstream of the dipoles prior to the transfer line, as shown in Fig.~\ref{fig:L4laserwirelayout}.

Other recent developments include the concept explored at J-PARC of a multi-laser-wire that uses confocal mirrors to reflected light through a set of parallel passes of the particle beam, thus avoiding the~need to physically scan the laserwire~\cite{bib:Miura2016}. Developments also include a longitudinal laserwire that is being developed to directly monitor the multi-dimensional phase space of H$^{-}$ bunches at the Front End Test Stand at the STFC Rutherford Appleton Laboratory, UK.~\cite{bib:Gibson2018}.

\subsection{Shintake monitor}
While electron laserwires are capable of monitoring the profile of micron-scale beams of around $100 \times 1\,\mu$m, the final focus of a future TeV e$^{+}$e$^{-}$ linear collider requires monitoring nanometer-scale beams of around 100\,nm by a few nm at the interaction point. This formidable challenge stimulated the proposal of an ingenious beam diagnostic technique by Tsumoru Shintake~\cite{bib:Shintake1992}.

The first Shintake Monitor was developed at the Final Focus Test Beam at SLAC~\cite{bib:Shintake1992, bib:Tenenbaum1995, bib:Balakin1995, bib:Shintake1999} which achieved and measured a vertical beam size of $\sigma_y 73$\,nm with approximately 10\% resolution. More recently a Shintake monitor was installed at ATF2~\cite{bib:Suehara2010, bib:Yamaguchi2010, bib:Yan2014} with the aim of measuring the vertical beam size to 37\,nm, scaled by energy from the ILC design of 5.9\,nm, and has demonstrated capability in beam size measurement with 5-10\% stability~\cite{bib:Yan2014, bib:YanPhD2014}. 

\begin{figure}[h!]
\begin{center}
\includegraphics[width=0.8\textwidth]{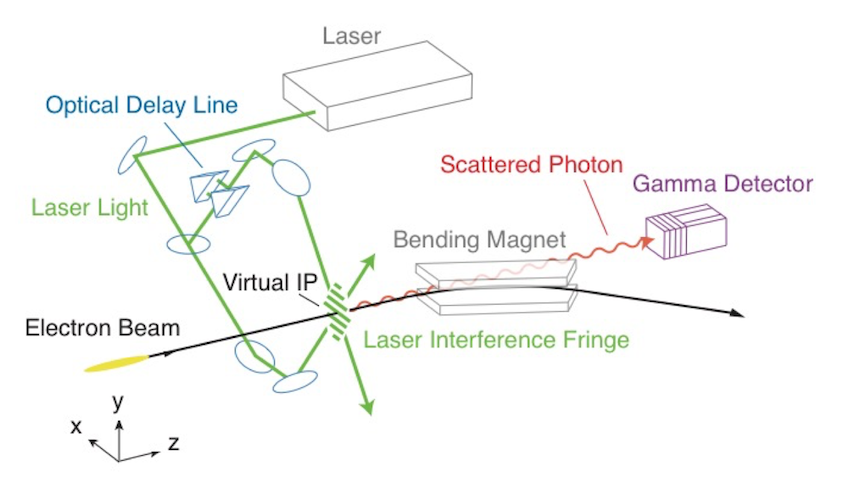} 
\caption{\small Layout of Shintake monitor~\cite{bib:Yamaguchi2010}}
\label{fig:ShintakeLayout}
\end{center}
\end{figure}

\begin{figure}[h!]
\begin{center}
\includegraphics[width=0.8\textwidth]{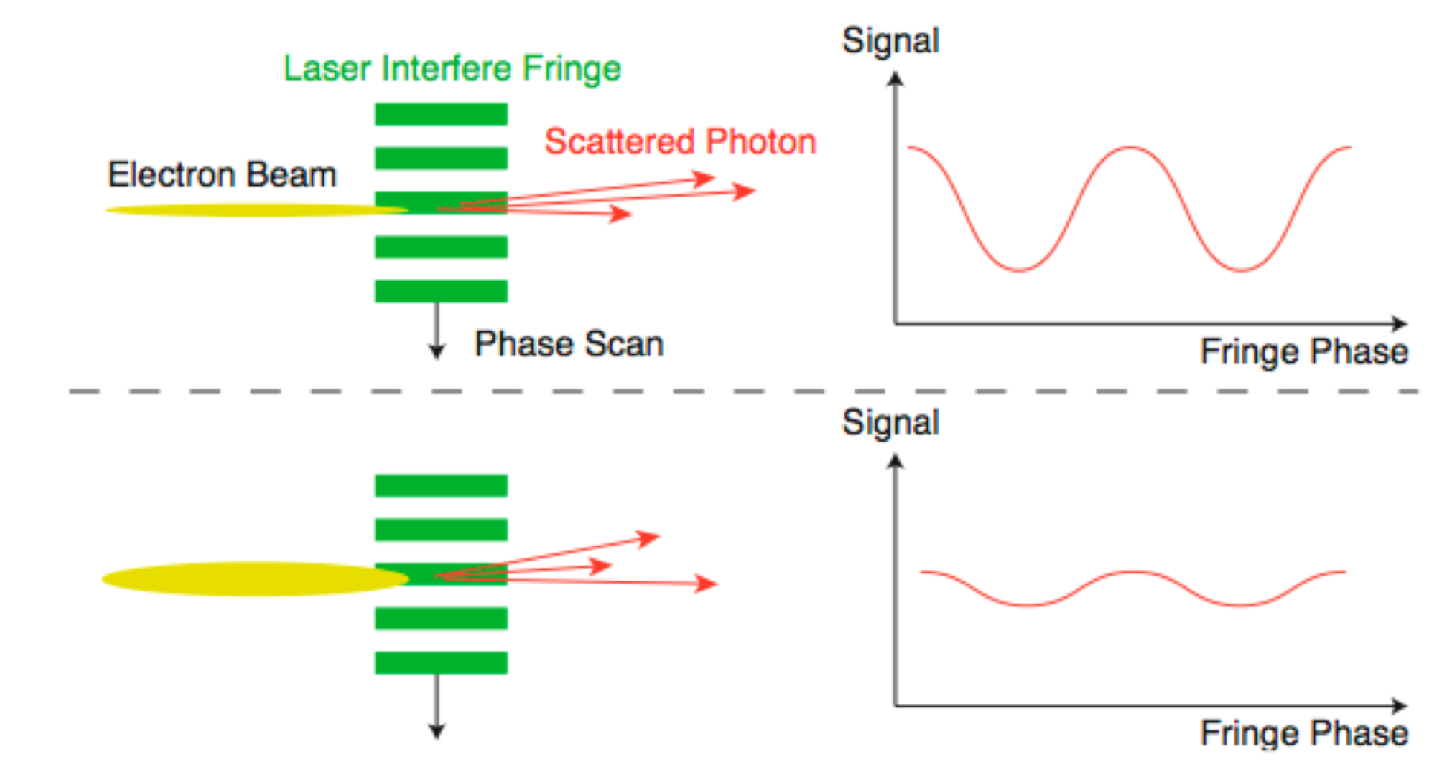} 
\caption{\small Principle of Shintake monitor, showing a change in the modulation depth for different beam sizes~\cite{bib:Yan2014}}
\label{fig:ShintakePrinciple}
\end{center}
\vspace{-10pt}
\end{figure}

The concept is similar to an electron laserwire based on inverse Compton scattering, as described in Section~\ref{sec:electronlaserwires}, except that two laser beams are made to overlap to create a pattern of interference fringes,  through which the electron beam passes, as shown in Fig.~\ref{fig:ShintakeLayout}.
The scattering rate is dependent on the relative size of the electron beam compared to the interference fringe pitch, and the relative transverse position, which can be controlled by scanning the relative fringe phase with respect to the~electron beam. 
The key measurement parameter is the modulation depth of the scattered photon intensity at the downstream detector, as the phase of the fringes is adjusted, as shown in Fig.~\ref{fig:ShintakePrinciple}.

\noindent The interference fringe pitch is
 \begin{equation}
 d = \frac{\lambda}{2 \sin{(\theta/2)}}\,,
\end{equation}
where $\theta$ is the crossing angle between the propagation directions of the two laser beams. Using a Nd:YAG Q-switched pulsed laser of SHG 532\,nm, a remote controlled reconfiguration of the crossing angle between $\theta=174^\circ$, $30^\circ$ and $2-8^\circ$ enables the measurable vertical beam size from 20\,nm to 6\,$\mu$m. The~performance is dependent on the any mismatch in angle between the electron beam and laser fringes (fringe tilt), and on the relative phase or equivalently position (phase or position jitter). Source of fast jitter have been studied and suppressed to achieve stability performance at the level of 6\%~\cite{bib:Yan2014}. 





\end{document}